\documentclass[aps,pre,twocolumn,superscriptaddress,showpacs,10pt]{revtex4-1}

\usepackage{graphicx}
\usepackage{mathtools}
\usepackage{amssymb,amsmath}
\usepackage[caption = false]{subfig}
\usepackage{booktabs}
\usepackage{placeins}
\usepackage{verbatim}
\usepackage{comment}
\usepackage{color}
\usepackage{float}
\usepackage{multirow}
\usepackage[utf8]{inputenc}
\usepackage[export]{adjustbox}

\newcommand{\ra}[1]{\renewcommand{\arraystretch}{#1}}

\newcommand{\reffig}[1]{Fig.~\ref{fig:#1}}
\newcommand{\refsec}[1]{Sect.~\ref{sec:#1}}
\newcommand{\reftable}[1]{Table~\ref{table:#1}}
\newcommand{\refeqs}[1]{equation~(\ref{eq:#1})}

\newcommand{\uv}[1]{{\,\bf \hat{#1}}}
\newcommand{\rsym}[0]{\mathcal{I}}

\newcommand{\etal}[0]{\textit{et al.}\textcolor{white}{a}}

\newcommand{\Rey}[1]{${Re}$}
\newcommand{\real}[1]{\mathcal{R}\{#1\}}

\newcommand{\EkD}[1]{EQ$_{#1}$}
\newcommand{\PkD}[1]{PO$_{#1}$}
\newcommand{\DCn}[1]{DC$_{#1}$}
\newcommand{\TkD}[1]{QP$_{#1}$}
\newcommand{\authedits}[1]{\textcolor{black}{#1}}

\definecolor{gray}{rgb}{0.5,0.5,0.5}

\begin{document}

\title{Heteroclinic and Homoclinic Connections in a Kolmogorov-Like Flow}

\date{\today}

\author{Balachandra Suri}
\affiliation{IST Austria, 3400 Klosterneuburg, Austria}
\author{Ravi Kumar Pallantla}
\affiliation{School of Physics, Georgia Institute of Technology, Atlanta, GA 30332, USA}
\author{Michael F. Schatz}
\affiliation{School of Physics, Georgia Institute of Technology, Atlanta, GA 30332, USA}
\author{Roman O. Grigoriev}
\affiliation{School of Physics, Georgia Institute of Technology, Atlanta, GA 30332, USA}

\begin{abstract}
Recent studies suggest that unstable recurrent solutions of the {Navier-Stokes equation provide new insights into dynamics of turbulent flows.}
In this study, we compute an extensive network of dynamical connections between such solutions in a weakly turbulent quasi-two-dimensional Kolmogorov flow that lies in the inversion-symmetric subspace.
In particular, we find numerous isolated heteroclinic connections between different types of solutions -- equilibria, periodic, and quasi-periodic orbits -- as well as continua of connections forming higher-dimensional connecting manifolds. We also compute a homoclinic connection of a periodic orbit and provide strong evidence that the associated homoclinic tangle forms the chaotic repeller that underpins transient turbulence in the symmetric subspace.

\end{abstract}


\maketitle

\section{Introduction}

Turbulent fluid flows are ubiquitous; they can be found in the atmosphere and the oceans, water and oil pipelines, and even in the human aorta.
Despite its great practical relevance, a tractable description of turbulent dynamics has remained elusive. 
However, recent  numerical \cite{kerswell_2005, kawahara_2012} and experimental \cite{hof_2004, lozar_2012, suri_2017a} studies have shown that {unstable recurrent} solutions  of the Navier Stokes equation, which governs fluid flows, may prove pivotal in solving this longstanding problem. 
Often termed Exact Coherent States (ECSs), such solutions exist for the same parameters as turbulence but are more amenable to numerical analysis given their simple (e.g., steady or periodic)  temporal behavior. 

\begin{figure}[!htbp]
\includegraphics[width=3.3in,valign=c]{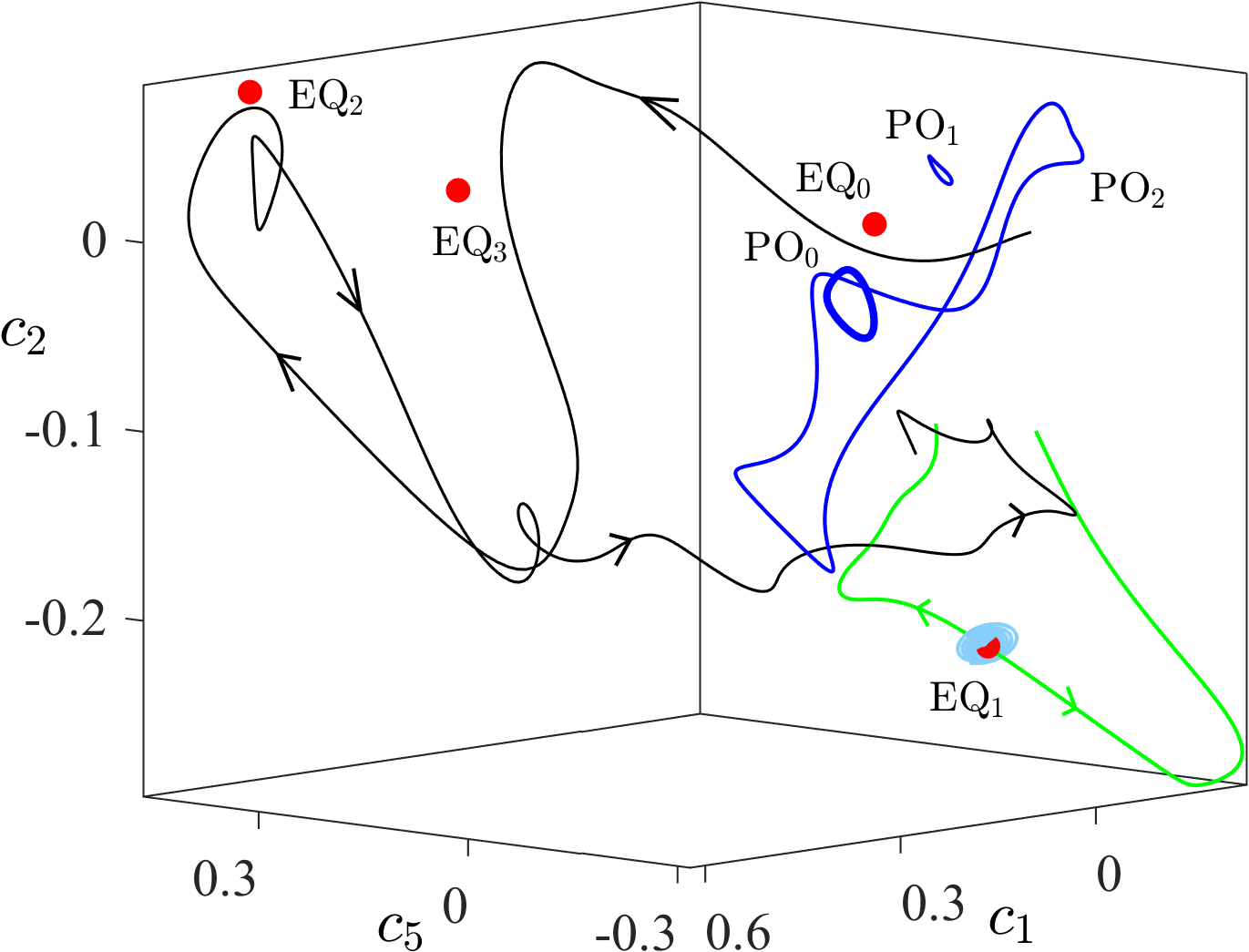}
\caption{\label{fig:all_sols} Low-dimensional projection of the state space generated using data from numerical simulation of a weakly turbulent Kolmogorov-like flow. Instantaneous snapshots of the flow in the physical space are associated with points in the state space. Turbulent evolution is represented by a trajectory (black curve) passing through neighborhoods of unstable equilibria (spheres) and periodic orbits (loops). Stable and unstable manifolds (green curves and light blue surface) of these solutions constrain the dynamics in their neighborhoods. \authedits{The projection method is described in Appendix \ref{sec:projections}.}} 
\end{figure}

The state space description of turbulence best illustrates the dynamical role of ECSs \cite{hopf_1948, gibson_2008}.
A turbulent flow in physical space maps to a winding trajectory in state space,  with each point on it representing a flow field (see supplementary videos 1-7) \footnote{}.  
In contrast, ECSs such as {steady} and time-periodic flows are simpler objects (fixed points, closed loops), as shown in  \reffig{all_sols}.
Being unstable, each ECS is a saddle in state space;  trajectories in its unstable manifold are repelled away, while those in the stable manifold converge to the ECS  \cite{duguet_2008b,viswanath_2009, suri_2017a, suri_2018}. 
Heteroclinic (homoclinic) trajectories -- which originate in the unstable manifold of an ECS and terminate in the stable manifold of another (the same) ECS -- connect different ECSs and create a chaotic saddle in state space \cite{halcrow_2009,duguet_2008b,farano_2018a}. 
 Dynamics on ECSs, their stable manifolds, and on homo/heteroclinic connections are asymptotically non-chaotic.
In this geometrical {picture}, turbulence {represents a deterministic} walk between neighborhoods of {different} ECSs, guided by {the corresponding} dynamical  connections \cite{kerswell_2007, suri_2017a}, {both homoclinic and heteroclinic.}

Substantial numerical evidence has emerged for the dynamical relevance of ECSs in recent years, mostly from research on three-dimensional (3D) wall-bounded shear flows, such as  plane-Couette  \cite{kawahara_2001,kawahara_2005,viswanath_2007,gibson_2009}, pipe \cite{faisst_2003,wedin_2004,kerswell_2007, pringle_2007}, and channel flows \cite{waleffe_2001,itano_2001,toh_2003}.
Direct numerical simulation (DNS) of flows in small, spatially periodic domains \cite{kim_1987,hamilton_1995} {suggests} that turbulence at moderate Reynolds numbers ($Re$)  
is organized around unstable {solutions such as} equilibria (EQ) \cite{wang_2007,gibson_2008, gibson_2009}, traveling waves (TW) \cite{waleffe_2001,faisst_2003, wedin_2004,pringle_2007}, periodic orbits (PO) \cite{kawahara_2001, viswanath_2007, duguet_2008a}, and relative periodic orbits (RPO) \cite{cvitanovic_2010,budanur_2017a}. 
Here, TWs (RPOs) are solutions that {correspond to steady (time-periodic) states in a reference frame moving} in the direction of a continuous symmetry (e.g., along the axis of a pipe). 
Flow fields resembling ECSs were also observed in a few laboratory experiments \cite{hof_2004, lozar_2012, lemoult_2014, suri_2017a}, which further validated their relevance in turbulence.

The geometry of chaotic saddle shaped by invariant  (stable/unstable) manifolds and dynamical connections between ECSs, however, remains under-explored.
In particular, the connectivity of different neighborhoods can be determined by generating a dense set of trajectories spanning the unstable manifold of an ECS and identifying which neighborhoods are subsequently visited by each trajectory \cite{kerswell_2007, gibson_2008}.
In general, dynamically relevant ECSs have several ({three of more})  unstable directions \cite{gibson_2009, kerswell_2007, budanur_2017a, farano_2018a, suri_2018}, which renders {this procedure} computationally expensive.
To circumvent this challenge numerical studies to date have analyzed dynamics confined to invariant subspaces (e.g., symmetry subspaces, laminar-turbulent boundary) which reduces the number of dynamically relevant ECSs as well as the dimensionality of their unstable manifolds. 
Using this technique, several previous studies \cite{kawahara_2001, toh_2003, kerswell_2007, duguet_2008a, duguet_2008b, budanur_2017b} identified trajectories that originate at ECSs with only one or two unstable directions and subsequently approach another ECS.
However, such trajectories were not proven to asymptotically converge to an ECS. Hence, they can only be regarded as likely signatures of dynamical connections.

Even within {invariant subspaces, very few dynamical connections between ECSs have been found.}    
Gibson \etal \cite{gibson_2008} and Halcrow \etal \cite{halcrow_2009} computed four 
heteroclinic connections between unstable EQs in plane-Couette flow (PCF).
Using low-dimensional projections of state space, the authors showed that {turbulent trajectories are  transiently} guided by these connections. 
The structural stability of these connections -- their robustness to small changes in $Re$ -- was also discussed using dimension counting arguments \cite{smale_1967}.
Two homoclinic orbits of a PO in {PCF} were computed by 
van Veen \etal \cite{vanveen_2011b} using a multi-shooting algorithm \cite{vanveen_2011a}. 
The authors suggested that dynamics along these connections resemble ``bursting'' phenomenon observed in turbulent boundary layers \cite{kline_1967}. 
Riols \etal \cite{riols_2013}, in a study of subcritical transition to magnetorotational dynamo in Keplerian shear flows, computed both homoclinic and heteroclinic connections between unstable POs. 
Pershin \etal \cite{pershin_2019} found a heteroclinic connection from an EQ to a nearby PO in PCF.
In both Riols \etal and Pershin \textit{et al.}, connections were computed very close to the saddle-node bifurcations leading to the formation of these invariant solutions. 
Recently, Budanur \etal \cite{budanur_2019} computed a homoclinic orbit of a spatially localized RPO in pipe flow and {suggested that the associated homoclinic tangle underlies transient turbulence in this flow.}
All the connections listed above originate (terminate) at ECSs with only one/two (one/zero) unstable directions, which facilitates {the use of simple shooting and bisection algorithms}. 
Recently, Farano \etal \cite{farano_2018a}  {showed that an adjoint-based method can be used to find connections between neighborhoods of} unstable EQs in {PCF where the originating (destination) EQs has more than two (one) unstable direction.}

Despite {these advances}, an extensive exploration of state space to detect signatures of connections {between dynamically dominant ECSs has} not been carried out yet. 
For instance,  some previous studies computed connections between ECSs of the same type (i.e., between EQs or between POs) \cite{gibson_2008, halcrow_2009, vanveen_2011b} while others reported only isolated connections between POs and EQs \cite{riols_2013,pershin_2019}.   
The chaotic saddle, however, is shaped by {ECSs of different types} and a complex network of dynamical connections between them. 
To address this shortcoming of previous studies, we report in this article a systematic and exhaustive exploration of {low-dimensional} unstable manifolds of ECSs to detect dynamical connections between various types on ECSs. 

The system we numerically study is the quasi-two-dimensional (Q2D) Kolmogorov-like flow in a shallow electrolyte layer driven by a horizontal, (spatially) near-sinusoidal body force. 
{Q2D flows are computationally more tractable than 3D flows since they can be accurately described using a 2D model}  \cite{bondarenko_1979, suri_2014}.
In fact, the {relative simplicity} of 2D DNS has prompted researchers in recent years to  carry out {the most} systematic exploration of ECSs in 2D turbulence \cite{chandler_2013, lucas_2014, farazmand_2016}. 
For instance, Chandler \etal \cite{chandler_2013} studied 2D Kolmogorov flow on a periodic domain {to test whether turbulent statistics can be reproduced using suitable averages over time-periodic solutions} \cite{cvitanovic_1988}.
More recently, Suri \etal \cite{suri_2017a, suri_2018} have validated the dynamical role of EQs and their unstable manifolds, for the first time in laboratory experiments, using Q2D Kolmogorov-like flow as the test bed. 2D DNS in these studies \cite{suri_2017a, suri_2018} {were} performed on a numerical domain with lateral dimensions and boundary conditions identical to those in the experiment, facilitating quantitative comparison between DNS and experiment.

This article is structured as follows: In \refsec{2dns} we discuss the 2D model for Q2D flows and its {numerical implementation}. In Sections \ref{sec:2dm_e2} -- \ref{sec:po1b} we explore invariant manifolds of various EQs and POs and {identify} heteroclinic and homoclinic connections. The stability of dynamical connections to small changes in  Reynolds number {is discussed in} \refsec{lim_cycle}. In \refsec{transient} we discuss {the relation between the dynamical connections and transient turbulence}. Finally, we summarize our findings and discuss their significance in \refsec{summary}.

{\section{Quasi-2D Kolmogorov-like Flow}}\label{sec:2dns}

The evolution of weakly turbulent flows in an electromagnetically driven shallow electrolyte layer is modeled {here} using a strictly 2D equation \citep{suri_2014}
\begin{equation}\label{eq:2dns_mod_ndim}
\partial_t{\bf u} + \beta {\bf u}\cdot\nabla{\bf u} = -\nabla p + \frac{1}{Re}\left(\nabla^2 {\bf u}- \gamma{\bf u}\right) + {\bf f},
\end{equation} 
which is derived by averaging 3D Navier Stokes equation {in the vertical ($z$) direction} \cite{suri_2014}. 
Here,  {$\nabla$ is the gradient operator restricted to the horizontal dimensions $x$, $y$, and ${\bf u}(x,y,t)$ is the velocity field at the electrolyte-air interface which is assumed to be incompressible ($\nabla\cdot{\bf u}=0$). This} is an accurate approximation for moderate  Reynolds numbers ($Re\lesssim 40$)  \cite{suri_2014}. 
${\bf f}$ is the {nondimensionalized} depth-averaged  horizontal forcing profile, and $p$ {plays the role of} the 2D kinematic pressure. 
In Q2D flows, the solid boundary beneath the fluid layer causes a vertical  gradient in the magnitude of the horizontal velocity. 
The prefactor $\beta\neq 1$  and linear friction $-\gamma {\bf u}$ model the effective change in inertia and the shear stress, respectively, due to this vertical gradient \cite{bondarenko_1979, suri_2014}.
For the flow in experiments detailed in Suri \etal \cite{suri_2017a},  which we numerically study in this article, $\beta = 0.83$ and $\gamma = 3.22$. 
We note {that} these values differ significantly from  $\beta=1$ and $\gamma=0$ corresponding to an unphysical strictly 2D flow typically studied in numerics \cite{chandler_2013,farazmand_2016}. 
The Reynolds number $Re$ {describes} the strength of electromagnetic forcing and serves as {the parameter that controls} the complexity of flow.
The dimensional form of \refeqs{2dns_mod_ndim} and analytical expressions for $\beta$, $\gamma$, and $Re$ as functions of experimental parameters are provided in references \cite{suri_2014} and \cite{tithof_2017}.

DNS of the flow was performed using a finite-difference code previously employed in studies \cite{suri_2017a, suri_2018} and \cite{tithof_2017}. Velocity and pressure fields
on a computational domain with lateral dimensions $(L_x,L_y) = (14,18)$ 
were spatially discretized using a 280$\times$360 staggered grid with spacing $\delta x = \delta y = 0.05$ between grid points.   
No-slip boundary conditions were imposed on the velocity field, and spatial derivatives were approximated using second-order central finite difference formulas.
Temporal integration of \refeqs{2dns_mod_ndim} was performed using the semi-implicit P2 projection scheme to enforce incompressibility of the velocity field at each time step \citep{armfield_1999}.  
The linear (nonlinear) terms in \refeqs{2dns_mod_ndim}  were discretized in time using second order implicit Crank-Nicolson (explicit Adams-Bashforth) method.
A time step $\delta t \leq 1/100$ was used for temporal integration to ensure the CFL number ${\max\{{u_x}\delta t/\delta x, u_y\delta t/\delta y\}} \leq 0.5$.  

In a  2D Kolmogorov flow, the forcing profile is strictly  sinusoidal, i.e., ${\bf f}\propto \sin(\pi y)\uv{x}$. 
In experiments detailed in  references \citep{suri_2017a, tithof_2017}, however, the electromagnetic forcing is sinusoidal only near the center of the domain and decays to zero at the boundaries. 
To replicate this forcing, we used a dipole lattice approximation  of the magnet array in the experiment and computed the resulting electromagnetic forcing. 
Comparison between experimentally measured forcing profile and the numerically estimated one was provided in Tithof \etal \cite{tithof_2017}.
The 2D forcing profile ${\bf f}$ computed from {the} dipole lattice model is anti-symmetric under the coordinate transformation  $\rsym:(x,y)\rightarrow(-x,-y)$, 
i.e., $\rsym{\bf f} = -{\bf f}$. This 2-fold inversion symmetry ($\rsym^2 = 1$) is  equivalent to rotation by $\pi$ about the $z-$axis passing through the lateral center of the computational domain. 
Under  $\rsym$, 
the velocity field transforms as $\rsym {\bf u}(x,y,t) \rightarrow -{\bf u}(-x,-y,t)$, which {makes} \refeqs{2dns_mod_ndim} equivariant under ${\rsym}$. 

\begin{figure*}[!ht]
\centering
\subfloat[]{\includegraphics[height=2.15in,valign=c]{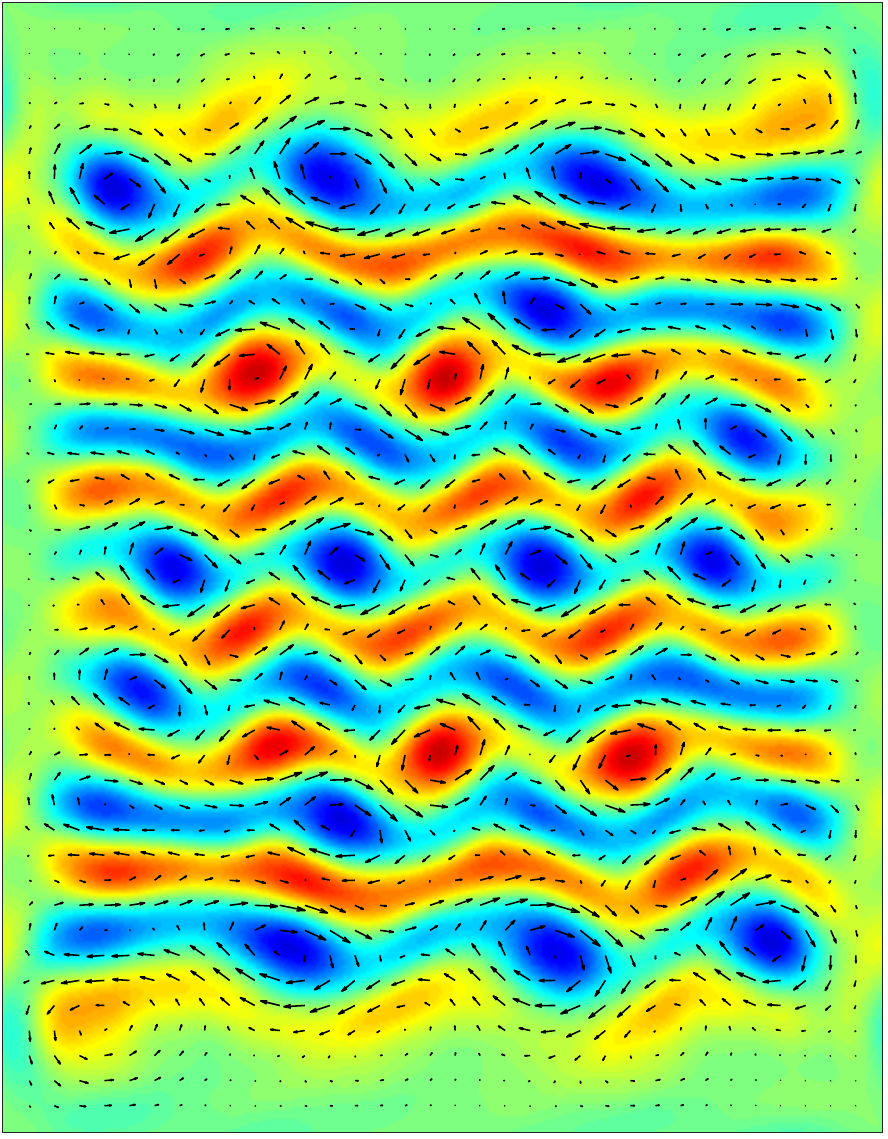}}\hspace{1mm}
\subfloat[]{\includegraphics[height=2.15in,valign=c]{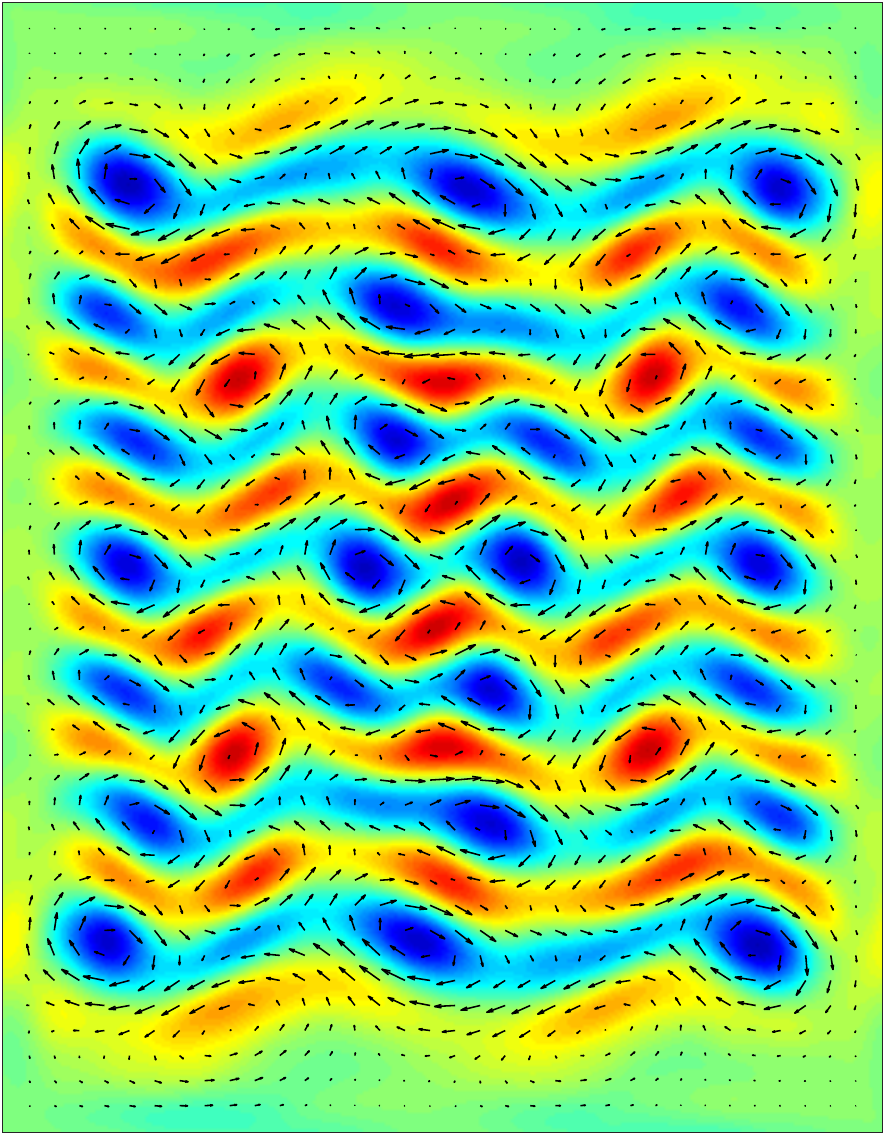}}\hspace{1mm}
\subfloat[]{\includegraphics[height=2.15in,valign=c]{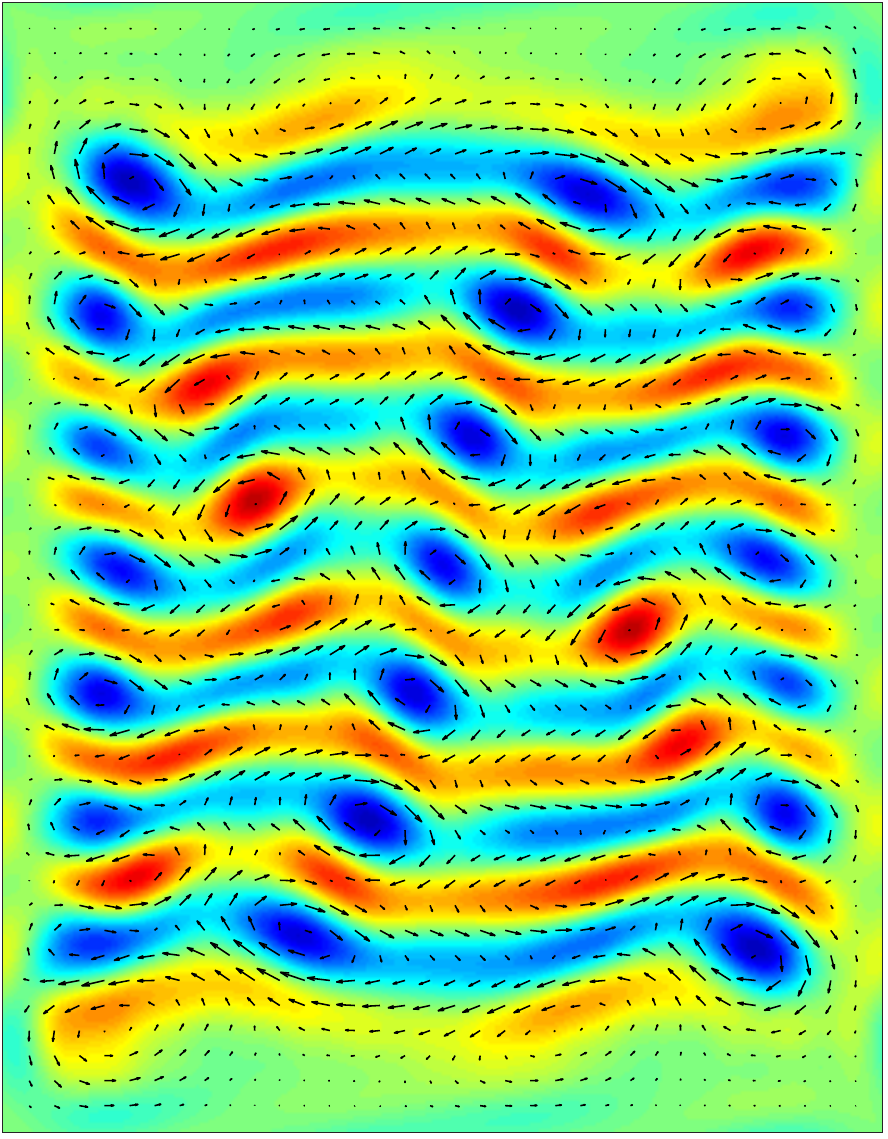}}\hspace{1mm}
\subfloat[]{\includegraphics[height=2.15in,valign=c]{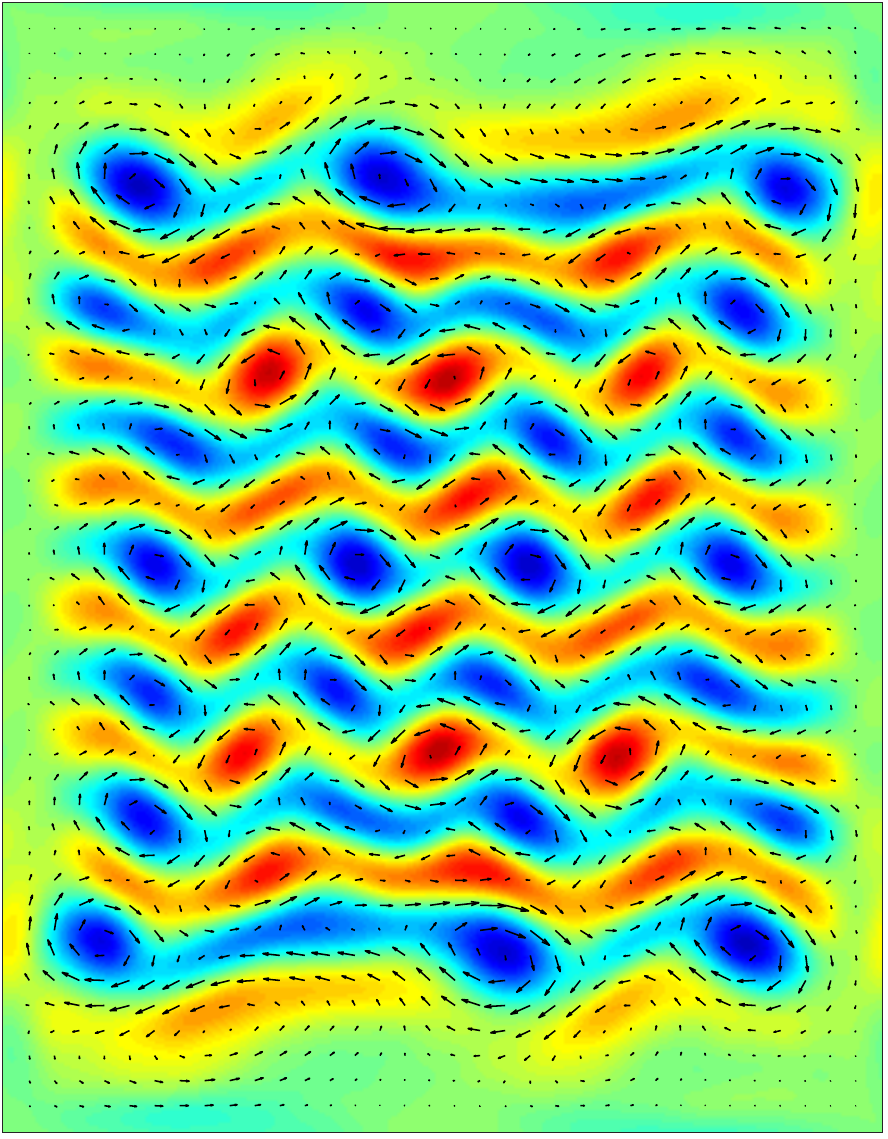}}
\caption{\label{fig:flow_fields} Rotationally symmetric equilibrium solutions (a) \EkD{2} (b) \EkD{0} (c) \EkD{1} (d) \EkD{3}. Color (black arrows) indicates vorticity (velocity).}
\end{figure*}

A consequence of this equivariance is that the symmetry of a rotationally invariant flow is preserved during its time evolution, i.e., {the symmetry subspace $\mathcal{S}=\{{\bf u}\in\mathcal{M}\,|\,\rsym {\bf u}={\bf u}\}$ of $\mathcal{M}$ is invariant. 
Here, $\mathcal M$ represents the full state space.
All ECSs and connections between them we report in this article lie in $\mathcal{S}$.
Since trajectories in $\mathcal{S}$ are generally unstable,  numerical errors will accumulate such that ${\bf u}(t)$ will eventually leave $\mathcal{S}$ even if ${\bf u}(0)\in\mathcal{S}$. To prevent this, we augmented the numerical integrator by projecting ${\bf u}(t)$ back into $\mathcal{S}$ after every time step \cite{suri_2017b}.
\vspace{1mm}
\section{Results}\label{sec:results}

{As $Re$ is increased, the Kolmogorov-like flow undergoes a number of bifurcations before transitioning to weak turbulence at $Re\approx 18$} \cite{tithof_2017}.
At $Re\approx22$ the flow in full state space {$\mathcal{M}$}
is chaotic, which was confirmed by computing the Lyapunov spectrum of a long turbulent trajectory using continuous Gram-Schmidt orthogonalization \cite{farmer_1983, suri_2018}.   
In this dynamical regime we previously identified 31 unstable equilibria of the 2D model \cite{suri_2018}. Twenty eight of these were {outside of $\mathcal{S}$}. Furthermore, their unstable manifolds were relatively high-dimensional  in  {$\mathcal{M}$}  (with the number of unstable directions as low as three and as high as twenty) and thus computationally forbidding to map out.
Hence, we instead started our analysis with the  equilibria in $\mathcal{S}$, {labeled \EkD{0}, \EkD{1}, and \EkD{2}}.

The stability properties of these equilibria and other recurrent solutions we found in $\mathcal{S}$ are summarized in \reftable{sol_summary}. We have listed the number of unstable directions $N_u$ and the leading eigenvalues $\lambda$  for EQs (Floquet exponents for POs) both in the symmetry invariant subspace $\mathcal{S}$ and in the rest of state space, i.e., $\mathcal{M}\backslash\mathcal{S}$.
All ECSs we computed have two or fewer unstable directions in $\mathcal{S}$, which allows their unstable manifolds to be well-approximated with a dense set of trajectories \cite{suri_2018}.
Note that \EkD{0} and \PkD{0} are stable in $\mathcal{S}$. Hence, turbulence in $\mathcal{S}$ can be transient, with  trajectories converging to one of these two ECSs asymptotically in time. This is indeed what numerical simulations show, as we discuss in \refsec{transient}.

\setlength{\tabcolsep}{5pt}
\begin{table}[!tbp]\centering
\ra{1.3}
\begin{tabular}{|c|c|c|c|c|}
\hline 
\multirow{2}*{}   & \multicolumn{2}{|c|}{$N_u$}     & \multicolumn{2}{|c|}{$\lambda$}  \\
\hline
&  $\ \mathcal{S}\ $ & $\mathcal{M}\backslash\mathcal{S}$ & $\mathcal{S}$ & $\mathcal{M}\backslash\mathcal{S}$ \\
\hline
{\EkD{0}} & 0 & 2 & $-0.004 \pm 0.020i$  &  $0.058 \pm 0.047i$\\
{\EkD{1}} & 1 & 2 & $0.017$              & $0.029$\\
{\EkD{2}} & 2 & 5 & $0.035 \pm 0.114i$  & $0.022$\\
{\EkD{3}} & 1 & 2 & $0.068$              & $ 0.029 \pm 0.029i$\\
{\PkD{0}} & 0 & 3 & $-0.003 \pm 0.059i$  & 0.151\\
{\PkD{1}} & 1 & 2 & 0.036                & $0.054 \pm 0.033i$\\
{\PkD{2}} & 1 & 3 & 0.044                & 0.049\\  
{\TkD{1}} & 1 & - & -                & -\\  \hline 
\end{tabular}
\caption{\label{table:sol_summary} {The number of unstable directions $N_u$ and the leading eigenvalue(s) (Floquet exponent(s)) $\lambda$ for the recurrent solutions we computed  in $\mathcal{S}$ at $Re=22.05$.}}
\end{table}

{Connection(s) from an origin (ECS$_{-\infty}$) to a destination (ECS$_{\infty}$) lie at the intersection of the unstable manifold of  ECS$_{-\infty}$ and the stable manifold of  ECS$_{\infty}$. Since the dimensionality of stable manifolds is very high ($O(10^5)$ in our case), we will search {the low-dimensional} unstable manifolds of ECSs for dynamical connections. 
Such unstable manifolds can be conveniently approximated by families of trajectories ${\bf u}({\bf p},t)$ that start near an ECS, where the unstable manifold is locally tangent to the linear subspace parametrized by the unstable and marginal eigenvectors of that ECS. Here, ${\bf p}$ parametrizes the family of manifold trajectories. 
The specifics of constructing a linear subspace, and consequently its parametrization, depend on the type of ECS (EQ or PO) as well as the dimensionality of its unstable manifold, {as discussed below}.

Table \ref{table:sol_summary} shows that several solutions we identified have a single unstable direction in $\mathcal{S}$.
The unstable manifold of such ECS is naturally divided into two halves, which correspond to the positive and negative halves of the corresponding linear subspace  (see \reffig{all_sols}). We shall hereafter refer to these halves as ${\bf u}^\pm({\bf p},t)$. 
Furthermore, ${\bf u}^\pm({\bf p},t)$ lie on the opposite sides of the stable manifold of the ECS,  which serves as a local separatrix and repels trajectories near the ECS along ${\bf u}^+({\bf p},t)$ or  ${\bf u}^-({\bf p},t)$. This allows us to compute connections terminating at ECSs with one unstable direction  using a simple bisection method \cite{itano_2001}.}
In the following sections we provide a detailed discussion of connections between various ECSs listed in \reftable{sol_summary}.

\subsection{Connections originating at \EkD{2}}\label{sec:2dm_e2}

{We start our analysis with equilibrium \EkD{2}, which has the  highest dimensional unstable manifold of all the EQs in $\mathcal{S}$. 
The vorticity field ($\omega = \nabla\times {\bf u}$) associated with \EkD{2} is shown in \reffig{flow_fields}(a).}  The 2D unstable manifold of \EkD{2} is locally tangent to the plane spanned by the complex conjugate pair of unstable eigenvectors $\uv{e}_1, {\uv e}_2 = \uv{e}_1^\ast$ associated with {$\lambda_{1,2} = \sigma\pm i \mu$, where $\sigma=0.035$ and $\mu=0.114$.}
To construct this 2D surface, {which lies entirely in $\mathcal{S}$}, we generated 360 initial conditions ${\bf u}(\theta,0)$ on a circle around \EkD{2} in the plane spanned by ${\uv e}_1$, ${\uv e}_2$:
\begin{equation}\label{eq:2dm_E2}
{\bf u}(\theta,0) = {\bf u}_{eq} + \epsilon\cos(\theta) \uv{e}_1'+\epsilon \sin(\theta)\uv{e}_2'.
\end{equation} 
Here, ${\bf u}_{eq}$ corresponds to \EkD{2} and $\uv{e}_1'$, $\uv{e}_2'$ are real orthonormal vectors constructed from the real and imaginary  parts of ${\uv e}_1$, $\uv{e}_2$. For numerical convenience we chose  $\epsilon = 0.002\cdot\|{\bf u}_{eq}\|$ and positioned the initial conditions ${\bf u}(\theta,0)$  at equal angular intervals $\Delta \theta = 2\pi/360$ on the circle. Numerical {integration of initial conditions} ${\bf u}(\theta,0)$ generates trajectories ${\bf u}(\theta,t)$ that {approximate the 2D manifold and parametrize it} by $\theta$, $t$. 
Each {manifold} trajectory ${\bf u}(\theta,t)$ was {computed on a temporal interval of length 25$\tau_c$, where $\tau_c = 12.5$ (nondimensional time units) is the average temporal auto-correlation time} \cite{suri_2018}.

\begin{figure}[!tp]
\centering
\includegraphics[width=3.3in]{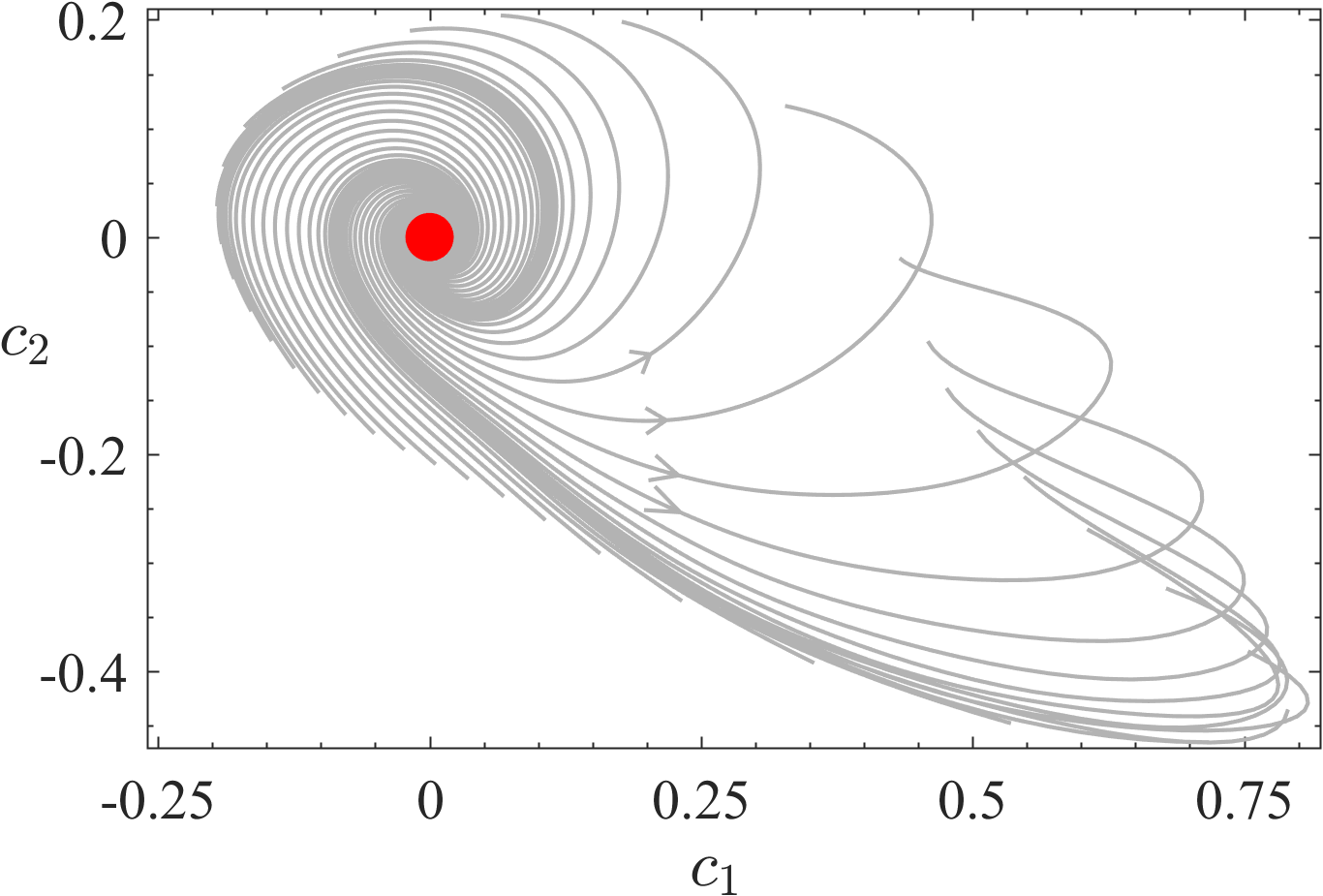}
\caption{\label{fig:2dm} {Trajectories ${\bf u}(\theta,t)$ (gray curves) approximating} the unstable manifold of \EkD{2} (red sphere). ${\bf u}(\theta,t)$ originate on a {small} circle around \EkD{2} that lies in a plane spanned by the complex conjugate pair of unstable eigenvectors $\uv{e}_1$ and $\uv{e}_2$. Coordinates $c_1, c_2$ are projections of ${\bf u}(\theta,t)$ onto orthonormal vectors constructed from $\uv{e}_1$, $\uv{e}_2$.}
\end{figure}

In the neighborhood of \EkD{2}, the various {trajectories} ${\bf u}(\theta,t)$ {initially} evolve spiraling outward, i.e., 
\begin{equation}\label{eq:spiral}
{\bf u}(\theta,t) \approx {\bf u}_{eq} + \epsilon e^{\sigma t} \cos(\theta +\mu t) \uv{e}_1' + \epsilon e^{\sigma t} \sin(\theta +\mu t)\uv{e}_2'.
\end{equation}
To illustrate this, we plotted a portion of the 2D unstable manifold in \reffig{2dm}, projected onto vectors $\uv{e}_1'$ and $\uv{e}_2'$. Only one in every ten trajectories generated is shown and the segment of each trajectory ${\bf u}(\theta,t)$ plotted corresponds to $0\leq t\leq11\tau_c$. 
Coordinates $c_1,c_2$ in \reffig{2dm} are the normalized inner products
\begin{equation}\label{eq:components}
c_{k}(t) = \frac{({\bf u}(\theta,t)-{\bf u}_{eq})\cdot\uv{e}'_k}{D_c},
\end{equation} 
where, $D_c = \max_{t,t'} \|{\bf u}(t)-{\bf u}(t')\| = 300$ is the empirically estimated maximum separation between two points on an 800$\tau_c$-long  turbulent trajectory in $\mathcal{S}$, \authedits{which defines the ``diameter'' of the chaotic repeller}. 
Normalizing distances with $D_c$ ensures that points separated by nearly unit distance in the low-dimensional projection are very far apart in full state space.
Notice that,  farther away from \EkD{2} ($|c_1|,|c_2|>0.2$) the {shape} of ${\bf u}(\theta,t)$ becomes fairly complicated due to the nonlinearity of the governing  \refeqs{2dns_mod_ndim}.
\begin{figure}[!tbp]
	\includegraphics[width=3.3in]{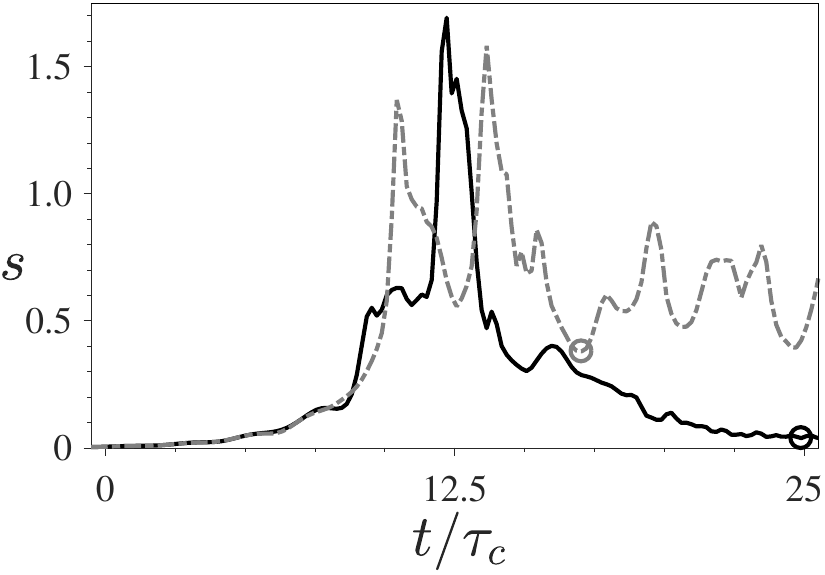}
	\caption{\label{fig:sss} State space speeds $s(\theta,t)$ along two  trajectories in the unstable manifold of \EkD{2} that exhibit contrasting dynamics. Solid (dashed) curve is speed along a trajectory that approaches (does not approach) an equilibrium closely after leaving the neighborhood of \EkD{2}. The lowest speed $s_m(\theta) = \min_{t}s(\theta, t)$ for $t/\tau_c \in [10,25]$ for each curve is marked using an open circle.} 
\end{figure}
It is not known \textit{a priori} which ${\bf u}(\theta,t)$  approach an ECS (EQ or PO) closely after leaving the neighborhood of \EkD{2}. 
Previously, Gibson \etal \cite{gibson_2008} and Halcrow \etal \cite{halcrow_2009}  inferred possible dynamical connections by inspecting low-dimensional state space projections. In contrast,  Riols \etal \cite{riols_2013} analyzed time-series of  (magnetic) energy and identified close passes to POs using intervals with  ``periodic'' behavior. Both these techniques, however, require visual inspection and detecting signatures of dynamical connections cannot be automated.  
In this study we tested two methods to detect signatures of dynamical connections {that do not require laborious visual inspection}. 
The method employed for detecting connections that terminate at EQs proved very effective and is discussed next. A 
method for detecting connections {to} POs using recurrence analysis \cite{duguet_2008b} is  discussed in Appendix \ref{sec:recurrence}.}

\begin{figure}[!tbp]
	{\includegraphics[width=3.3in]{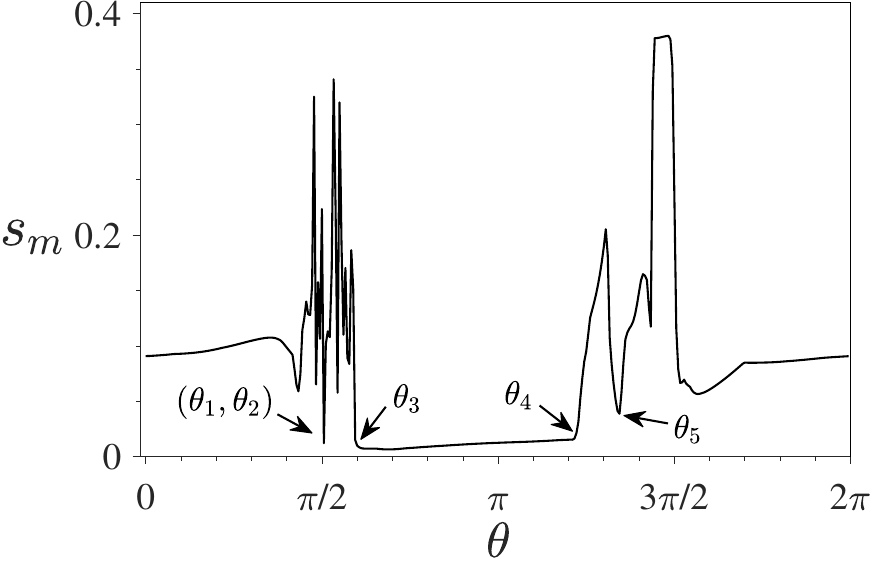}}
	\caption{\label{fig:s_vs_theta} Minimum state space speed $s_m(\theta) = \min_t s(\theta,t)$ along each trajectory ${\bf u}(\theta,t)$ in the 2D unstable manifold of \EkD{2}. Trajectories corresponding to $s_m \ll 1$ are possible dynamical connections from \EkD{2} to an equilibrium.} 
\end{figure}

For each trajectory  ${\bf u}(\theta,t)$ in the unstable manifold of \EkD{2}, we computed the normalized instantaneous state space speed $s(\theta,t)$, which we define as \cite{suri_2017a, suri_2018}:
\begin{equation}
s(\theta,t)  = \frac{\tau_c}{\|{\bf  u}(\theta,t)\|}\left\|\partial_t {\bf u}(\theta,t)\right\|.
\end{equation}
Since {$\partial_t {\bf u}=0$ ($s=0$) for any EQ, $s(\theta,t)\ll 1$} indicates that a  trajectory ${\bf u}(\theta,t)$ in state space is near an EQ \cite{suri_2017a,acharya_2017}. 
In the physical space, this corresponds to the evolution of {the} flow dramatically slowing down. 
Figure \ref{fig:sss} shows, as examples,  the speed plots for two different manifold trajectories  ${\bf u}(\pi/2,t)$ and ${\bf u}(3\pi/2,t)$. {Clearly, ${\bf u}(\theta,t)$ lie in the linear neighborhood of \EkD{2} for $0< t \lesssim 10\tau_c$ ($s\ll 1$) and $s(\theta,t)$ grows exponentially according to} \refeqs{spiral}. 
After  this initial transient, however, dynamics {described by} various trajectories can be {qualitatively} very different.
For instance, the {shape of} $s(3\pi/2,t)$  (gray curve) suggests that  ${\bf u}(3\pi/2,t)$ displays turbulent evolution.  
In contrast, $s(\pi/2,t)$ (black curve) suggests that ${\bf u}({\pi/2},t)$  approaches an EQ closely after a brief turbulent excursion.
\begin{figure*}[!tbp]
	\centering
	\subfloat[]{\includegraphics[height=2.9in,valign=c]{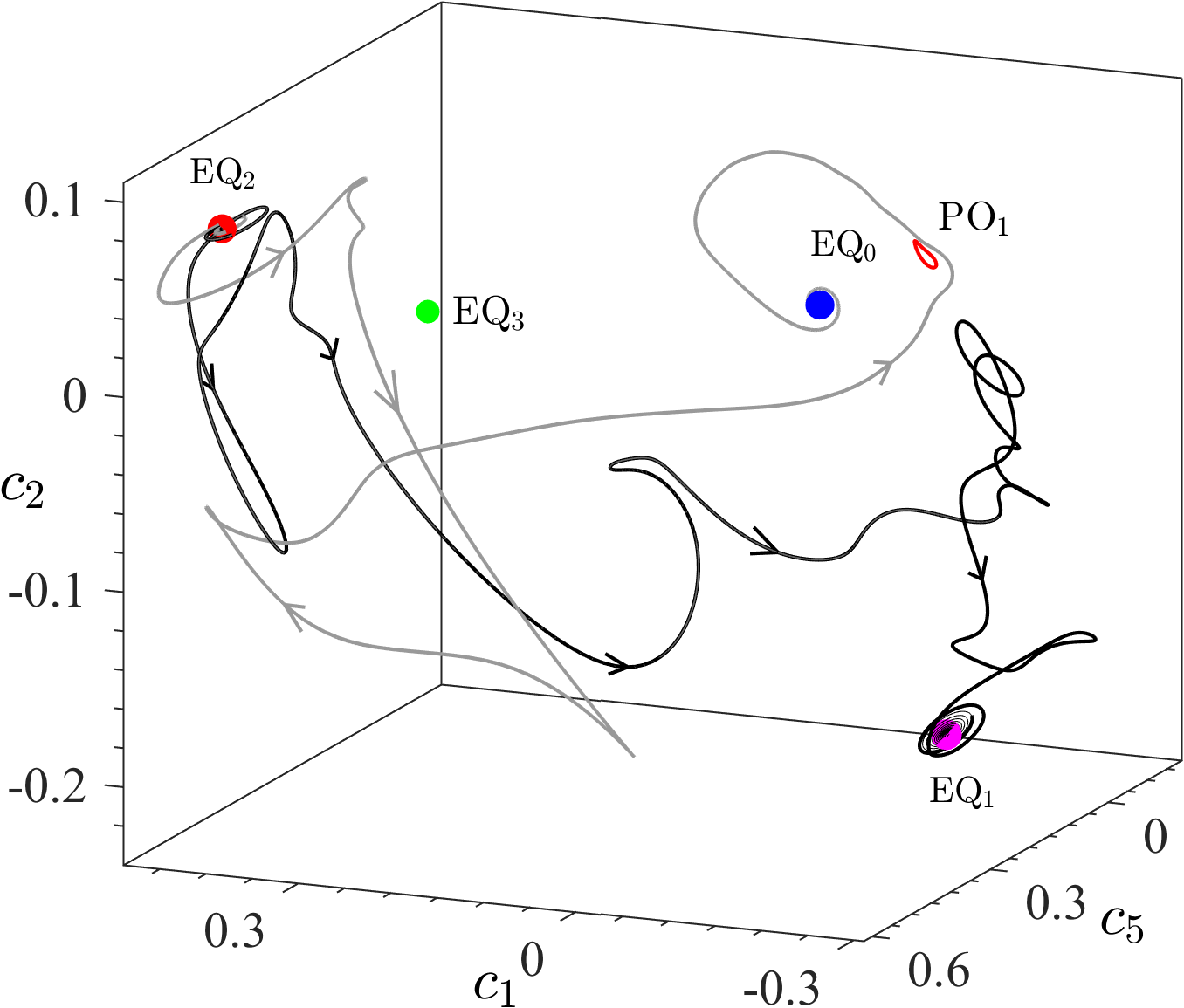}}\hspace{1mm}
	\subfloat[]{\includegraphics[height=2.9in,valign=c]{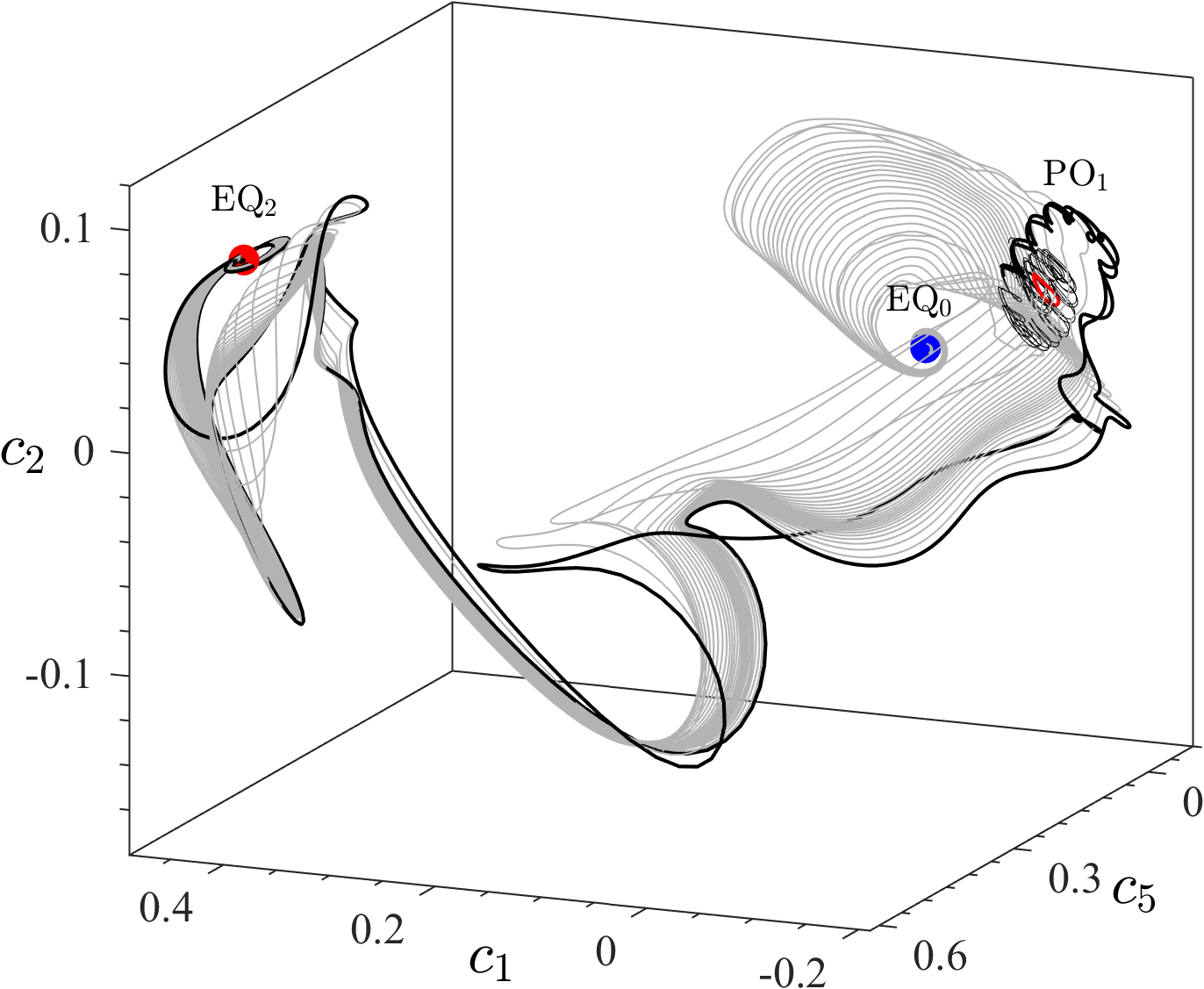}}
	\caption{\label{fig:hetero_E2D} Heteroclinic connections from \EkD{2}  to various ECSs. (a) A connection (gray curve) to \EkD{0} that belongs to the family \DCn{1} and an isolated connection \DCn{3} (black curve) to \EkD{1} (b)  Family of connections \DCn{2} (gray curves) that terminate at \EkD{0} and isolated connections \DCn{6}, \DCn{7} (black curves) to \PkD{1}. 
	} 
\end{figure*}
Hence, to test if a trajectory ${\bf u}(\theta,t)$ approaches an EQ \textit{after} it departs from the neighborhood of \EkD{2}, we computed the minimum speed $s_m(\theta) = \min_t s(\theta,t)$ for $10\tau_c \leq t \leq 25\tau_c$, {which is shown in \reffig{s_vs_theta}.} 
The low values of $s_m(\theta)$ for $\theta \in \left(\theta_1,\theta_2\right)$, $\theta\in\left (\theta_3,\theta_4\right)$, and $\theta={\theta_5}$ strongly suggest that the corresponding trajectories ${\bf u}(\theta,t)$ closely shadow either heteroclinic or homoclinic connections from \EkD{2} to other EQs.
In \reffig{s_vs_theta}, the uncertainty in $\theta$ is limited by the angular resolution $\Delta \theta$ in the initial conditions ${\bf u}(\theta,0)$. 
Consequently, to compute exact dynamical connections, the estimates for $\theta$ may require refinement in some cases. 
The labels $\theta_1$ through $\theta_5$ in \reffig{s_vs_theta} correspond to these refined values.
Lastly, $s_m(\theta)> 0$ even for trajectories ${\bf u}(\theta,t)$ that correspond exactly to dynamical connections. This is because, minima in the $s(\theta,t)$ are computed on a finite temporal interval, while dynamical connections converge ($s_m = 0$) to the destination EQs only in infinite time.

The destination EQs that different dynamical connections approach were computed using a Newton solver \cite{saad_1986, kelley_2003, mitchell_2013} initialized with the flow field ${\bf u}(\theta,t)$ at the instant when $s(\theta,t) = s_m(\theta)$. 
{For $\theta\in\left(\theta_1,\theta_2\right)$ and $\theta\in(\theta_3,\theta_4)$ the solver converged to \EkD{0}, which is a stable node in $\mathcal{S}$. For $\theta = \theta_5$ the solver converged to \EkD{1}, which is a saddle with one unstable direction in $\mathcal{S}$. 
The flow fields corresponding to \EkD{0} and \EkD{1} are shown in Figs. \ref{fig:flow_fields}(b) and \ref{fig:flow_fields}(c). 
Dimensional arguments (discussed in \refsec{lim_cycle}) suggest that connections from \EkD{2} to \EkD{0} should comprise a one-parameter family forming a two-dimensional manifold, just like those in the interval $(\theta_3,\theta_4)$. However,  the difference $\theta_2-\theta_1$ is smaller than the resolution $\Delta\theta$ in case of the interval $(\theta_1,\theta_2)$. Consequently, $\theta_1$,  $\theta_2$ cannot be distinguished in \reffig{s_vs_theta}. 
Hereafter, we will refer to $\theta$ in this narrow interval $(\theta_1,\theta_2)$ collectively as $\theta_n$. The wide  interval $(\theta_3,\theta_4)$ will be referred to as $\theta_w$.}
Lastly, to confirm that the dynamical connections originating at \EkD{2} indeed terminate at either \EkD{0} or \EkD{1}, we need to make sure that the distance 
\begin{equation}\label{eq:D0}
D_0 = \min\limits_t\frac{\|{\bf u}(\theta,t)-{\bf u}_{eq}\|}{D_c},
\end{equation}
vanishes in the limit $t\to\infty$. Here,  ${\bf u}_{eq}$ is the velocity field corresponding to the destination EQ. For all connections reported in this study, we chose $D_0<0.005$ as the criterion for convergence.

Let us start our analysis with the trajectories ${\bf u}(\theta_n,t)$ and ${\bf u}(\theta_w,t)$ that approach the equilibrium \EkD{0} which is stable in $\mathcal{S}$.
The eigenvalues of weakly contracting modes of \EkD{0} are  $\lambda_{1,2} = -0.004 \pm 0.020i$, $\lambda_{3,4}=-0.012 \pm 0.42i$.
Consequently, most nearby trajectories in $\mathcal{S}$ converge to \EkD{0} at a rate determined by the real part of $\lambda_1$ (${\real{\lambda_1}}$). Hence, we {computed trajectories ${\bf u}(\theta_n,t)$ and ${\bf u}(\theta_w,t)$ for an interval of duration  $350\tau_c \gg 1/\real{\lambda_{1}} \approx 10\tau_c$. 
At the end of numerical integration both ${\bf u}(\theta_n,t)$ and ${\bf u}(\theta_w,t)$ approached \EkD{0} to within a distance  $D_0<10^{-8}$, confirming that we have indeed found heteroclinic connections from \EkD{2} to \EkD{0}. 
Hereafter, we shall refer to these dynamical (heteroclinic) connections corresponding to  ${\bf u}(\theta_n,t)$ and ${\bf u}(\theta_w,t)$ as \DCn{1} and \DCn{2}, respectively.
\begin{figure*}
	\centering
	\includegraphics[width=6.7in]{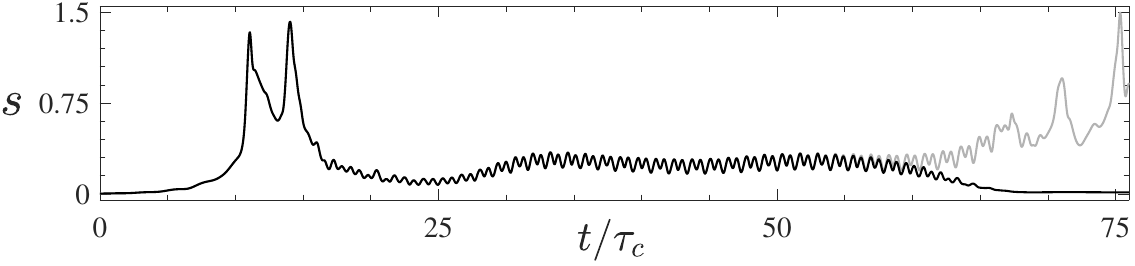}
	\caption{\label{fig:bisection_sss} State space speed along trajectories ${\bf u}(\theta_4^-, t)$ (black) and ${\bf u}(\theta_4^+,t)$ (gray) that separate after shadowing a periodic orbit.} 
\end{figure*}
A connection from the family \DCn{1} is shown (gray curve) in \reffig{hetero_E2D}(a) and thirty connections 
from \DCn{2} that are equally spaced on the interval $(\theta_3,\theta_4)$  are shown in \reffig{hetero_E2D}(b). 
Both figures show the projection of state space onto an orthogonal basis constructed from the stable eigenvectors $\uv{e}_1$, $\uv{e}_2$, and $\uv{e}_5$ of \EkD{0}. 
\authedits{The vectors $\uv{e}_1$, $\uv{e}_2$ were chosen to visualize asymptotic dynamics along the connections terminating at \EkD{0}. The vector $\uv{e}_5$ was chosen because trajectories  ${\bf u}(\theta,t)$ far away from \EkD{0} have large components along {this direction}.  We note that, unless mentioned otherwise,  these modes are used for all  state space projections in this article.}
  
In \reffig{hetero_E2D}, both \DCn{1} and \DCn{2}  converge to \EkD{0} spiraling inwards, almost entirely confined to the $c_1-c_2$ plane. This is a consequence of the large separation between the real values of $\lambda_{1,2}$ and $\lambda_{3,4}$  which makes {$\uv{e}_1$, $\uv{e}_2$} the only dynamically relevant eigenvector pair near \EkD{0}. 
Another interesting feature of \reffig{hetero_E2D}(b) is that \DCn{2} initially forms a flat strip bounded by manifold trajectories ${\bf u}(\theta_3,t)$ and ${\bf u}(\theta_4,t)$ (black curves). 
However, farther away from \EkD{2},  this strip widens and folds such that ${\bf u}(\theta_3,t)$ and ${\bf u}(\theta_4,t)$ come {close to each other, trace loops which are strikingly similar in shape, and eventually merge (see inset in \reffig{projection_bisection}). 
This is not an artifact} of low-dimensional projection of the state space. As we explain in the following section, {the black trajectories in \reffig{hetero_E2D}(b) correspond to  heteroclinic connections from \EkD{2} to a different ECS (\PkD{1})}.

Unlike {\DCn{1}} and \DCn{2}, the trajectory ${\bf u}(\theta_5,t)$ approaches the solution \EkD{1} which has one unstable direction (cf. \reftable{sol_summary}).
Since the stable/unstable manifolds of \EkD{1} guide the evolution of trajectories in its neighborhood, we explored their geometry to compute the connection from \EkD{2} to \EkD{1}. Trajectories 
{${\bf u}(\theta,t)$} for $\theta \approx \theta_5$ that approach \EkD{1} should subsequently depart following its unstable manifold, which {coincides with a pair of trajectories}
\begin{equation}\label{eq:1dm}
{\bf u}^\pm(t) = {\bf u}_{eq} \pm \epsilon e^{\lambda_1 t}\uv{e}_1,
\end{equation}
in the linear neighborhood of \EkD{1} (cf. \reffig{all_sols}). Here, $\uv{e}_1$ is the unstable eigenvector of \EkD{1} and {$\epsilon>0$ is some small constant.
In particular, we found {a pair of adjacent trajectories ${\bf u}(\theta,t)$ and ${\bf u}(\theta+\Delta\theta,t)$ with $\theta\approx\theta_5$}
approach \EkD{1} and  depart its neighborhood in opposite directions shadowing ${\bf u}^+(t)$ and ${\bf u}^-(t)$, respectively. 
Hence, a heteroclinic connection \DCn{3} from \EkD{2} to \EkD{1} lies between these two trajectories, i.e., in the stable manifold of \EkD{1}}.

{The projection of \DCn{3} (black curve) is shown in \reffig{hetero_E2D}(a). 
It was computed by refining the estimate for $\theta_5$ to within $\delta\theta=2^{-12}\Delta\theta$ using bisection  \cite{itano_2001,gibson_2008, halcrow_2009}.}
This refinement reduced the separation between ${\bf u}(\theta_5,t)$ and \EkD{1} to $D_0 = 0.002$  (cf. \refeqs{D0}). 
Further refinement in $\theta$ did not significantly reduce $D_0$ since numerical noise on ${\bf u}(\theta_5,t)$ amplifies rapidly in the direction of unstable eigenvector  $\pm\uv{e}_1$ of \EkD{1}.
\authedits{This behavior stems from the strong asymmetry in the real parts of unstable ($\lambda_1 = 0.017$) and weakly {stable} ($\lambda_{2,3}= -0.003 \pm 0.109i$) eigenvalues of \EkD{1}.} 
{Nevertheless, a better accuracy can be achieved} by employing multi-shooting \cite{vanveen_2011a} or approximating \DCn{3} as a piece-wise continuous solution \cite{toh_2003}. Using the latter technique {allowed us to decrease $D_0$ to less than $10^{-6}$. 	
{Lastly, equation \eqref{eq:1dm} governs the evolution of trajectories in the unstable manifold of \EkD{1} only in its linear neighborhood. 
Farther away, both ${\bf u}^\pm(t)$ display turbulent evolution for $t\gg100\tau_c$ and eventually approach \EkD{0}. 
Hence these two trajectories define (long) heteroclinic connections (\DCn{4}, \DCn{5}) from \EkD{1} to \EkD{0}. }	

\subsection{Dynamics Near \PkD{1} and  \EkD{3}} \label{sec:separatrix}
While trajectories ${\bf u}(\theta,t)$ for $\theta \in (\theta_3,\theta_4)$ quickly converge to \EkD{0}, trajectories just outside this interval exhibit qualitatively different dynamics;  this can be inferred from $s_m$ changing abruptly at $\theta_3$ and $\theta_4$.
This qualitative difference suggests that {trajectories ${\bf u}(\theta_3,t)$ and ${\bf u}(\theta_4,t)$ play the role of separatrices on the unstable manifold of \EkD{2} and hence lie} in the stable manifold of another ECS. 
In case of ${\bf u}(\theta_4,t)$, for example, we found this ECS  by inspecting the two trajectories ${\bf u}(\theta_4^-,t)$ and ${\bf u}(\theta_4^+,t)$  obtained as a result of successive bisections; here  $\theta_4^\pm=\theta_4\pm\delta\theta$ with $\delta\theta = 2^{-12}\Delta\theta$. 
The  corresponding state space speed plots are shown in \reffig{bisection_sss}, which suggest that ${\bf u}(\theta_4^\pm,t)$ evolve almost indistinguishably for about  $55\tau_c$ and subsequently  separate. 
For $t/\tau_c\in[30,55]$ the state space speed oscillates with a period of approximately $1.4\tau_c$, suggesting that  ${\bf u}(\theta_4^\pm,t)$ shadow a periodic orbit during this interval.

Using the flow field corresponding to  ${\bf u}(\theta_4^+,t)$ at  $t/\tau_c = 55$ as an initial condition into a Newton solver, we indeed found an unstable periodic orbit \PkD{1} with a period $T=1.32\tau_c$. 
A similar refinement using bisection showed that trajectories {at $\theta_3^\pm = \theta_3\pm\delta\theta$} also approach \PkD{1} and separate after shadowing it for an extended {period of time} (the corresponding speed plots are not shown). 
Figure \ref{fig:projection_bisection} shows the state space projection of ${\bf u}(\theta_4^\pm,t)$ {and} ${\bf u}(\theta_3^\pm,t)$ approaching \PkD{1}, shadowing it closely (inset), and subsequently leaving its neighborhood. The projection coordinates here are the same as in \reffig{hetero_E2D}, but the viewing angle is different. 
Lastly, the result that ${\bf u}(\theta_3^\pm,t)$ and ${\bf u}(\theta_4^\pm,t)$ approach \PkD{1} is consistent with the folding of \DCn{2} shown in \reffig{hetero_E2D}(b).

\PkD{1} has a single real unstable direction {in $\mathcal{S}$} with an associated Floquet exponent $\lambda_1 = 0.036$ (cf. \reftable{sol_summary}). 
Hence, the departure of ${\bf u}(\theta_3^\pm,t)$ and ${\bf u}(\theta_4^\pm,t)$ from {the neighborhood of} \PkD{1} is guided by its unstable manifold, which is {two-dimensional since it is associated with a PO which is itself one-dimensional}. This 2D unstable manifold can be constructed by evolving initial conditions  \cite{riols_2013}
\begin{equation}\label{eq:1dm_upo}
{\bf u}^{\pm}(\eta,0) = {\bf u}_{po} \pm \epsilon e^{\lambda_1\eta} \uv{e}_1,
\end{equation}
where ${\bf u}_{po}$ is a reference point on \PkD{1}, $\uv{e}_1$ is the Floquet vector at ${\bf u}_{po}$, {$\epsilon>0$ is sufficiently small (we chose $\epsilon= 0.001\cdot \|{\bf u}_{po}\|$), and
$\eta \in (0,T]$} parametrizes {different} initial conditions. 
A total of $720$ initial conditions, equally spaced in $\eta$, were generated and the corresponding trajectories ${\bf u}^\pm(\eta,t)$ were computed on an interval of length $35\tau_c$ to {approximate the manifold.}
We found that trajectories ${\bf u}^+(\eta,t)$  uneventfully converge to \EkD{0} while ${\bf u}^-(\eta,t)$ display turbulent evolution after leaving the neighborhood of \PkD{1}. 
To illustrate this, we plotted in \reffig{projection_bisection} a pair of trajectories  ${\bf u}^\pm(\eta,t)$ (red dashed curves)  evolving in opposite directions. 

\begin{figure}[!tp]
\vspace{1.25cm}
\includegraphics[width=3.3in]{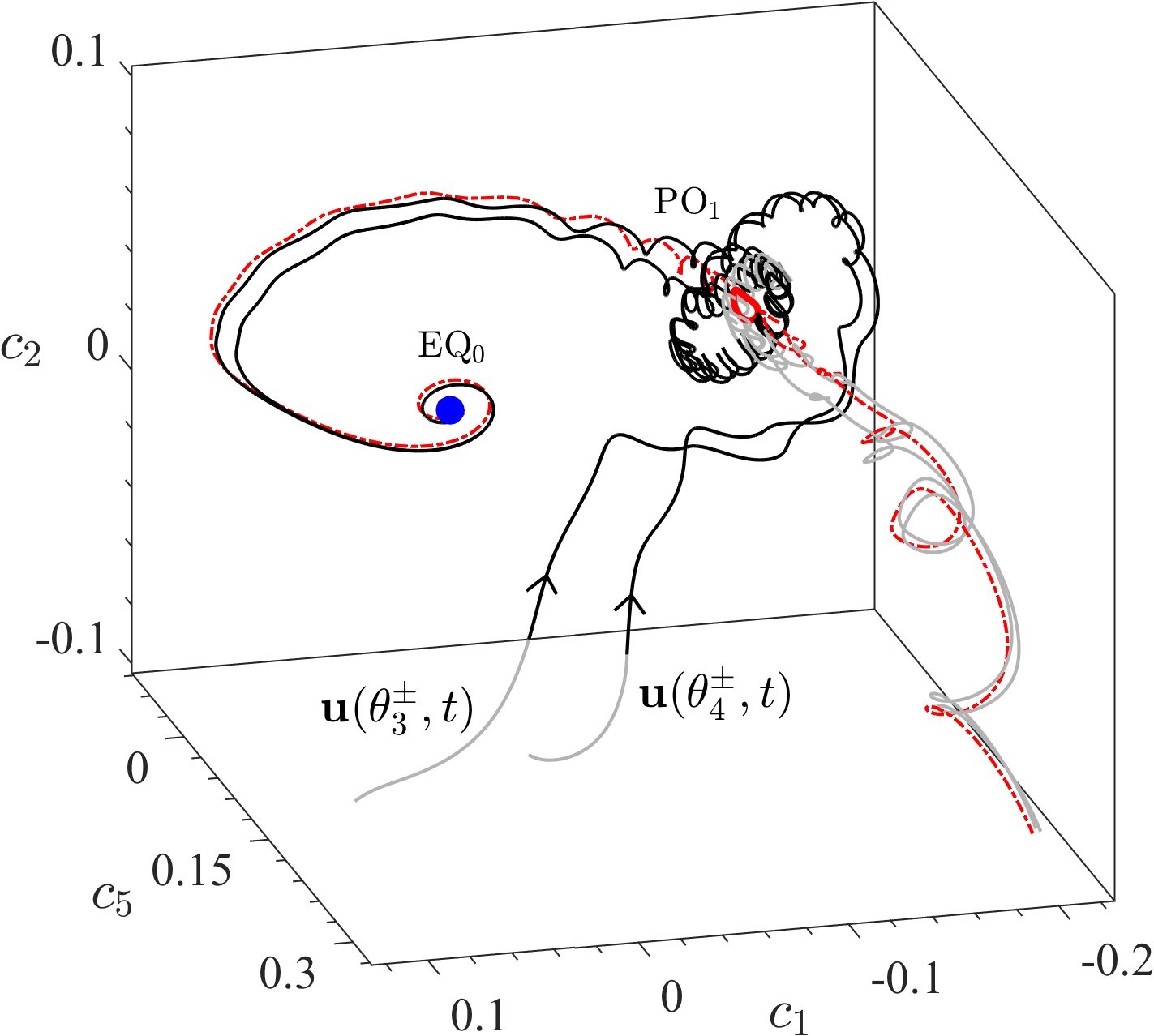}%
\begin{picture}(0,0)
\put(-83,186){\fbox{\includegraphics[width=2.5cm]{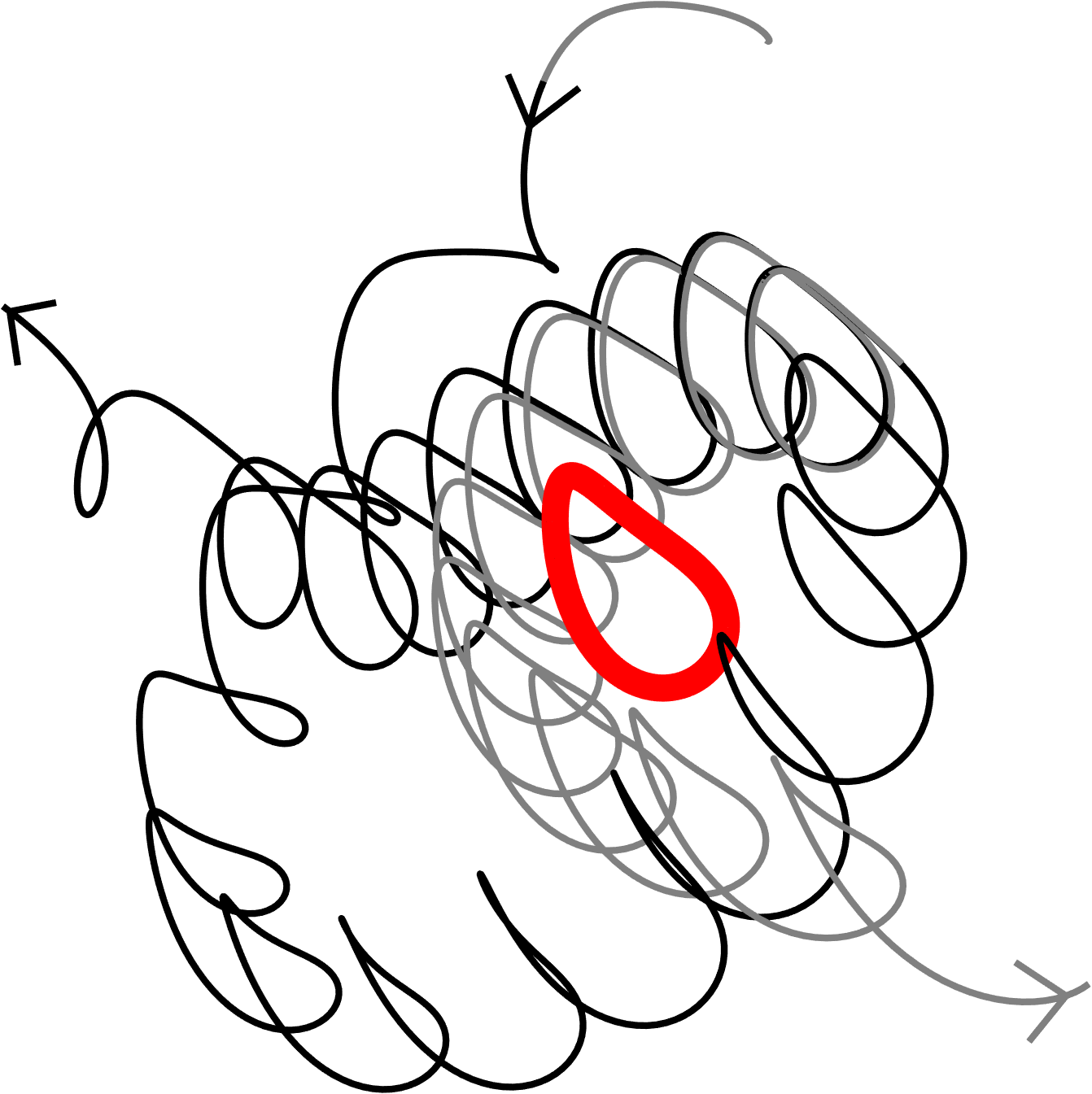}}}
\end{picture}
\caption{Geometry of state space around \PkD{1}. Trajectories ${\bf u}(\theta_4^\pm,t)$ approach \PkD{1} indistinguishably, but  depart from its neighborhood  in opposite directions (inset) {guided by the manifold trajectories ${\bf u}^\pm(\eta,t)$ (dashed red curves)}. ${\bf u}(\theta_4^-,t)$ (black curve) converges to \EkD{0} and ${\bf u}(\theta_4^+,t)$ (gray curve) becomes turbulent. Similar behavior is {manifested by} ${\bf u}(\theta_3^\mp,t)$.}
\label{fig:projection_bisection}
\end{figure}

Since \PkD{1} has a single unstable direction, its stable manifold divides the state space in the neighborhood of  \PkD{1} into two halves \cite{avila_2013}. 
We found that trajectories ${\bf u}(\theta_3^+,t)$ and ${\bf u}(\theta_4^-,t)$ smoothly converge to \EkD{0} after approaching \PkD{1} {and hence lie on one (the same) side of the stable manifold. Meanwhile} ${\bf u}(\theta_3^-,t)$ and ${\bf u}(\theta_4^+,t)$ display turbulent excursions {and hence lie on the opposite side of the stable manifold, as shown in \reffig{projection_bisection}.
Therefore,} there exist two heteroclinic connections from \EkD{2} to \PkD{1} sandwiched {between} ${\bf u}(\theta_3^\pm,t)$ (\DCn{6}) and ${\bf u}(\theta_4^\pm,t)$ (\DCn{7}), that lie on the {stable manifold} and asymptotically converge to \PkD{1}. 
Just like in the case of \DCn{3}, simple shooting cannot be employed to compute \DCn{6} and \DCn{7} in their entirety since they are {very} unstable.
Nevertheless, for sufficient refinement in $\theta$, the trajectories ${\bf u}(\theta_3^\pm,t)$ and  ${\bf u}(\theta_4^\pm,t)$ approximate \DCn{6} and \DCn{7}, respectively, reasonably accurately (cf. \reffig{hetero_E2D}(b)).
This was tested by computing the smallest distance  between ${\bf u}(\theta,t)$ and \PkD{1}
\begin{equation}\label{eq:D1}
D_1 = \min\limits_{t,\eta}\frac{\|{\bf u}(\theta,t)-{\bf u}_{po}(\eta)\|}{D_c},
\end{equation}
where ${\theta} = \theta_3^\pm$ or $\theta_4^\pm$ and ${\bf u}_{po}(\eta)$ are states along \PkD{1} parametrized by $\eta\in(0,T]$. We found that $D_1<0.005$ in both cases. 

\begin{figure}[!tp]
\centering
\includegraphics[width=3.3in]{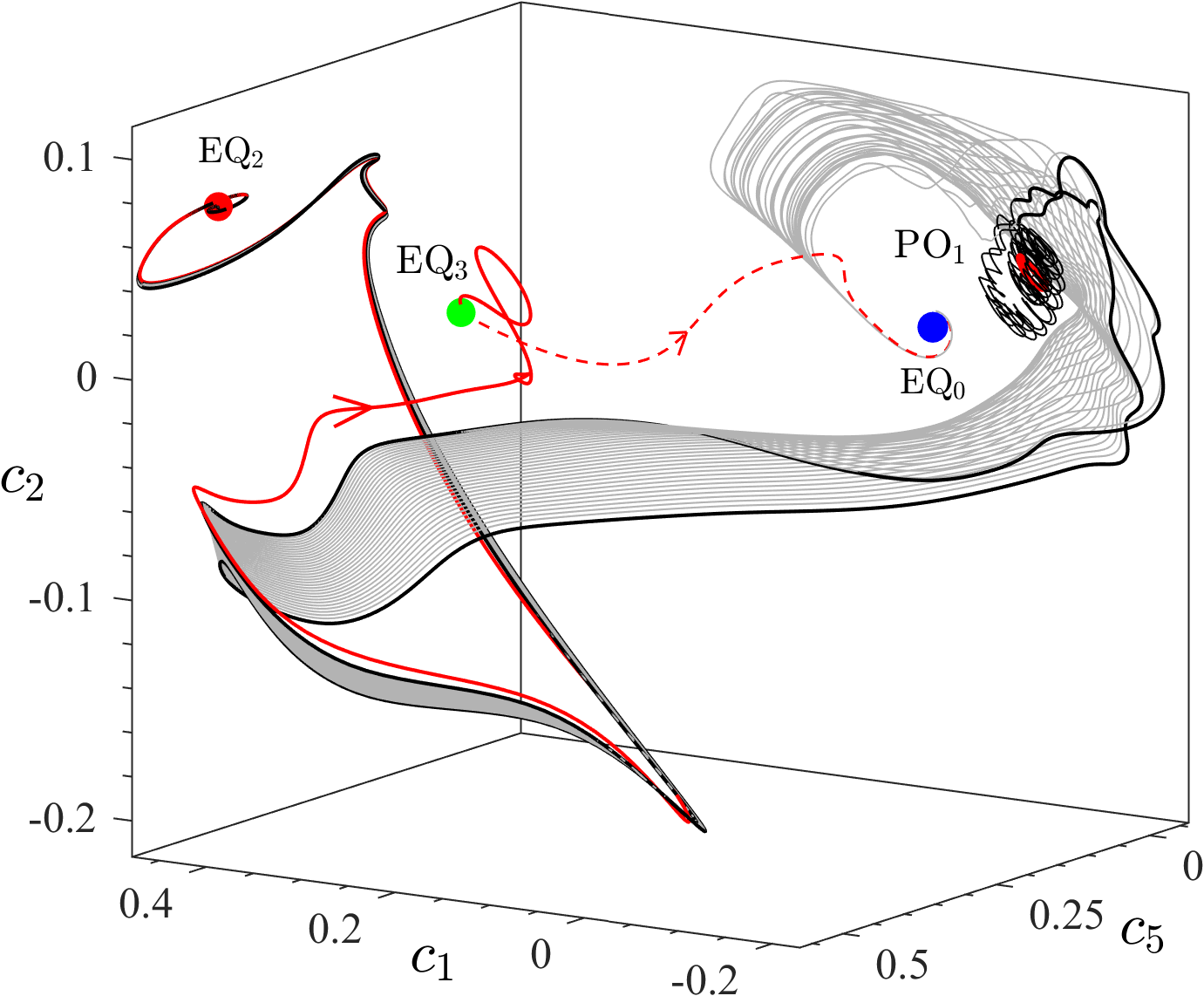}	
\caption{\label{fig:edge_piby2} Heteroclinic connections \DCn{8}, \DCn{9} (black curves) from \EkD{2} to \PkD{1} that sandwich the family of connections  \DCn{1} (gray curves) from \EkD{2} to \EkD{0}.  
The solid (dashed) 
red curve is the connection \DCn{10} (\DCn{11}) from \EkD{2} to \EkD{3} (\EkD{3} to \EkD{0}).} 
\end{figure}

\reffig{projection_bisection} shows a toroidal structure traced out by  ${\bf u}(\theta_4^\pm,t)$ and ${\bf u}(\theta_3^\pm,t)$ as these trajectories approach \PkD{1} along its stable manifold. A magnified view of this region shown in the inset illustrates the complicated shape of the manifold.
The numerous small loops correspond to the fast constant amplitude oscillation along \PkD{1}. In contrast, the large spiral corresponds to a slowly decaying oscillation described by the stable Floquet vectors $\uv{e}_2$, $\uv{e}_3$ of \PkD{1} with exponents $\lambda_{2,3} = -0.0044 \pm 0.0264i$. The characteristic time associated with this slow oscillation is $2\pi/0.0264\approx 20\tau_c$ and manifests itself in the weak modulation of the state space speed during the interval $t \in[30\tau_c, 50\tau_c]$  in \reffig{bisection_sss}. 

As mentioned earlier,  based on dimensional arguments,  \DCn{1} should be a {2D manifold which corresponds to a one-parameter family of trajectories connecting \EkD{2} to \EkD{0}. 
Since the corresponding interval $\theta_n$ is very narrow, $\theta_2-\theta_1<\Delta\theta$, we resampled a wider interval of width $2\Delta\theta$ that includes $\theta_n$ using 100 equally spaced initial conditions. 
This refinement showed that  $\theta_2-\theta_1\approx0.5\Delta\theta$ and all trajectories inside this interval converge to \EkD{0}, i.e., \DCn{1} is indeed a 2D manifold.}
Thirty trajectories from \DCn{1} are shown (as gray curves) in \reffig{edge_piby2}.  
The projection coordinates and the viewing angle are the same as in \reffig{hetero_E2D}. 
As in the case of \DCn{2}, the trajectories ${\bf u}(\theta_1,t)$ and ${\bf u}(\theta_2,t)$ at the left and right edges of \DCn{1} are separatrices on the 2D unstable manifold of \EkD{2}.
Using bisection, we identified that both ${\bf u}(\theta_1,t)$ and ${\bf u}(\theta_2,t)$ approach \PkD{1} closely ($D_1 < 0.005 $) and well-approximate heteroclinic connections \DCn{8}, \DCn{9} from \EkD{2} to \PkD{1} shown (as black curves) in \reffig{edge_piby2}. 

The refinement in initial conditions around $\theta_n$ also revealed that a few trajectories ${\bf u}(\theta,t)$  for $\theta_1-\Delta\theta<\theta<\theta_1$ approach an unstable equilibrium \EkD{3}. The flow field corresponding to \EkD{3} is shown in \reffig{flow_fields}(d). 
\EkD{3} has a single unstable direction in $\mathcal{S}$ and its unstable manifold is one-dimensional, similar to \EkD{1} (see \reffig{all_sols}).
Also, we found that a pair of adjacent trajectories from \EkD{2} approach \EkD{3} and subsequently depart in opposite directions. Hence, using bisection  we computed the heteroclinic connection \DCn{10} from \EkD{2} to \EkD{3} (solid red curve in \reffig{edge_piby2}).   
Additionally, we also found that  one of the manifold trajectories of \EkD{3} uneventfully converges to \EkD{0}. This connection (\DCn{11}) is also shown (as dashed red curve) in \reffig{edge_piby2}. 
The manifold trajectory from \EkD{3} evolving in the opposite direction, however,  also converged to \EkD{0} after a long turbulent excursion and hence corresponds to another connection (\DCn{12}).
\authedits{This qualitative difference in dynamics on {the two sides} of the stable manifold of \EkD{3} is not observed in case of  \EkD{1}. As mentioned {previously},  both trajectories in  the 1D unstable manifold of \EkD{1} display long turbulent excursions.}

\subsection{Connections originating at \PkD{1}} \label{sec:1dm_upo}
\begin{figure}[!tbp]
\centering
{\includegraphics[width=3.3in,valign=c]{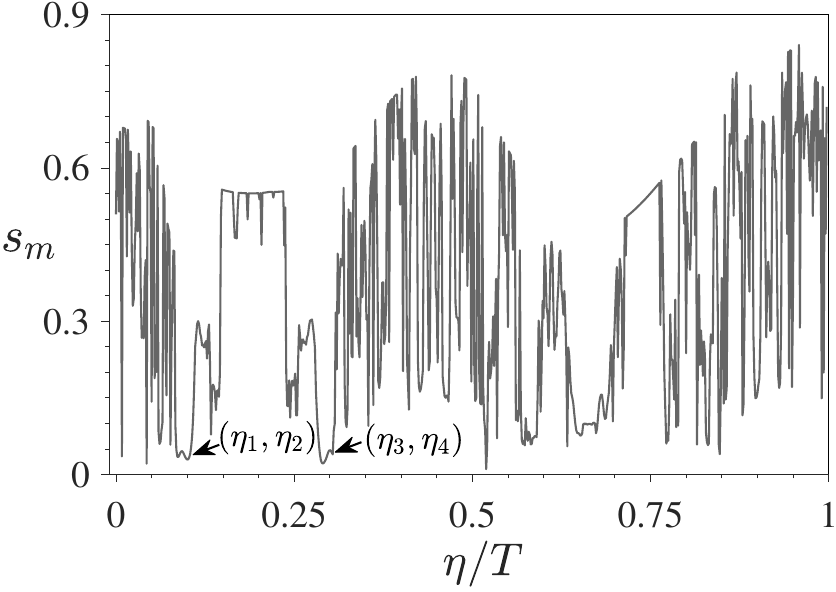}}
\caption{\label{fig:s_vs_eta} Minimum state space speed (black curve) $s_m$  along trajectories ${\bf u}^-(\eta,t)$ in the 1D unstable manifold of \PkD{1}. $\eta \in[0,T)$ parametrizes points along \PkD{1}.}
\end{figure}

As we have established, all trajectories ${\bf u}^+(\eta,t)$ 
from \PkD{1} quickly converge to \EkD{0} (cf. \reffig{projection_bisection}). 
These trajectories constitute a one-parameter family of heteroclinic connections from \PkD{1} to \EkD{0} which form a 2D manifold \DCn{13}. 
{The union of \DCn{6} and \DCn{7} forms a 1D boundary of the union of the 2D manifolds \DCn{2} and \DCn{13}. 
Similarly, the union of \DCn{8} and \DCn{9} forms a 1D boundary of the union of 2D manifolds \DCn{1} and \DCn{13}.}

Unlike ${\bf u}^+(\eta,t)$, the trajectories ${\bf u}^-(\eta,t)$ display turbulent excursions after departing from the neighborhood of \PkD{1}. 
Hence, we tested whether for some $\eta$ they approach other ECSs, as in the case of trajectories ${\bf u}(\theta,t)$ in the 2D unstable manifold of \EkD{2} (cf. \refsec{2dm_e2}). 
Since ${\bf u}^-(\eta,t)$ is also a one-parameter family of trajectories, we followed the procedure discussed in \refsec{2dm_e2} to search for signatures of dynamical connections originating at \PkD{1}.

To begin, we computed  the state space speed $s(\eta,t)$ for each trajectory ${\bf u}^-(\eta,t)$ (e.g., \reffig{rec}  in Appendix \ref{sec:recurrence}) and calculated $s_m(\eta) = \min_t s(\eta,t)$ for $t/\tau_c \in[{15,35}]$. 
From the plot of $s_m$ versus $\eta$, shown in \reffig{s_vs_eta}, we identified about 20 trajectories with $s_m \ll 1$ as possible dynamical connections. Surprisingly, \textit{all} these trajectories  converged to \EkD{0}. 
This was tested by extending the corresponding ${\bf u}^-(\eta,t)$ until the separation from \EkD{0} decreased to $D_0<0.01$. 
Since a detailed analysis of dynamics along all of these trajectories is not feasible, we {limit the discussion to} the families of connections {\DCn{14} that correspond to} the interval $(\eta_1,\eta_2)$ marked in \reffig{s_vs_eta}. 

\begin{figure}[!tp]
\centering
{\includegraphics[width=3.3in,valign=c]{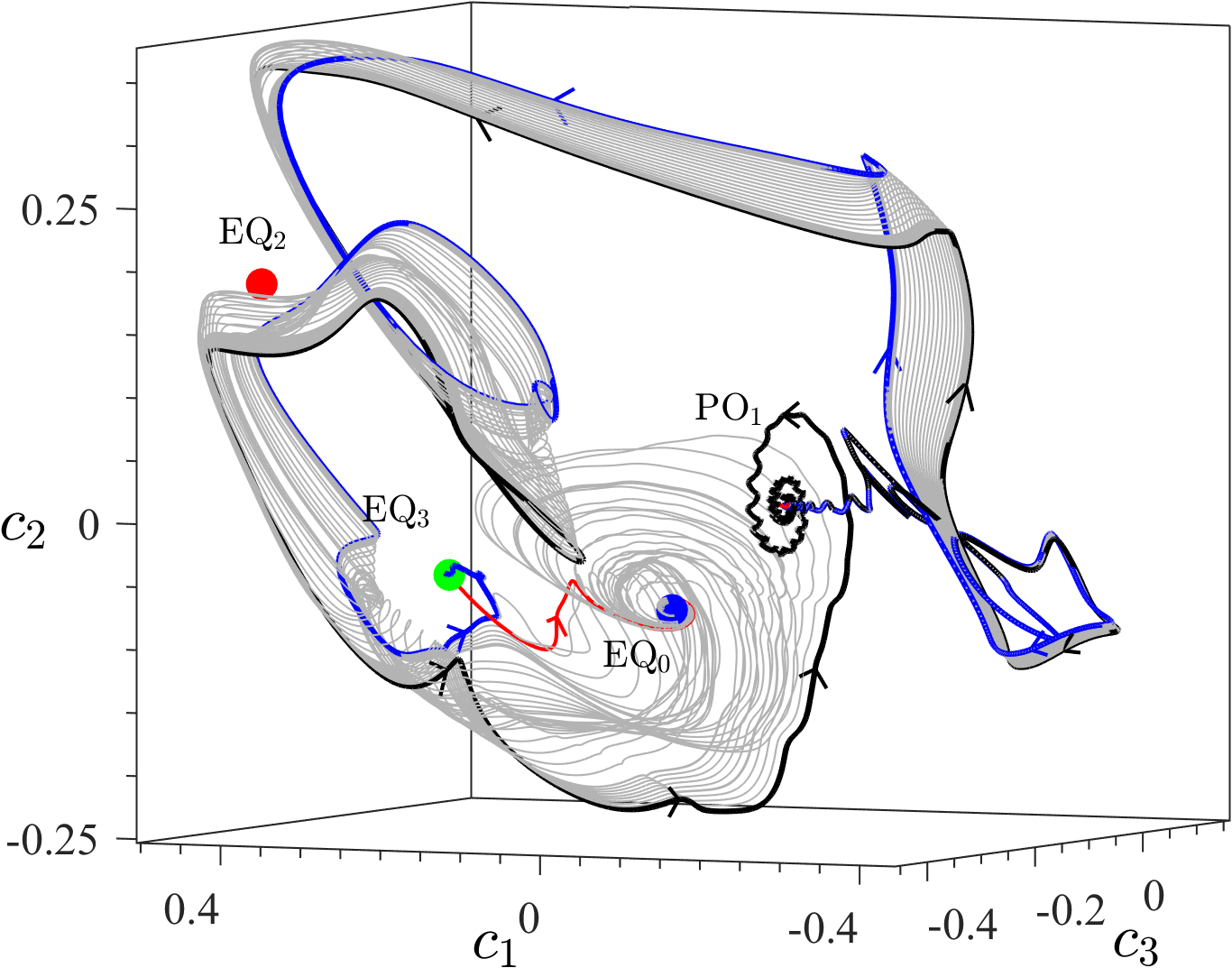}}
\caption{\label{fig:homoclinic} Family of heteroclinic connections (\DCn{14}, gray curves) from \PkD{1} to \EkD{0}. The black curve is a homoclinic orbit of \PkD{1} (\DCn{15}) and the blue curve is a heteroclinic connections from \PkD{1} to \EkD{3} (\DCn{16}). The red curve is \DCn{11} from \EkD{3} to \EkD{0}. 
}
\end{figure}

About thirty trajectories (gray curves) from the family \DCn{14} are shown in \reffig{homoclinic} which, \authedits{unlike all other figures,} employs a projection of the state space onto an orthonormal basis constructed from Floquet vectors $\uv{e}_1$ (unstable, real) and $\uv{e}_{2},\, \uv{e}_3$ (stable, complex conjugate pair) of \PkD{1}.  
\authedits{Trajectories in \DCn{14} leave the neighborhood of \PkD{1} along ${\bf e}_1$.}
After a brief turbulent excursion, they visit the neighborhood of \EkD{2} and finally converge to \EkD{0}. 
An interesting feature {of} \reffig{homoclinic} is that several trajectories in \DCn{14} appear to visit the neighborhood of \PkD{1} enroute to \EkD{0}.
This raises the question {of whether} a homoclinic orbit ${\bf u}^-(\eta,t)$ of \PkD{1} exists {near} the interval $(\eta_1,\eta_2)$. 
Since \PkD{1} has one unstable direction, a homoclinic orbit {should} be sandwiched between trajectories that show qualitatively different dynamics \cite{budanur_2019}. 
Such a behavior is {indeed} observed for trajectories near $\eta_1$ {and} $\eta_2$ where $s_m$ changes abruptly. 
Using bisection, we refined the estimate for $\eta_2$ and found that ${\bf u}^-(\eta_2,t)$ indeed approaches \PkD{1} very closely (within $D_1 < 0.005$) and therefore well-approximates a homoclinic connection \DCn{15} shown (black curve) in \reffig{homoclinic}.
\authedits{The projection used in this figure allows visualization of the asymptotic dynamics along \DCn{15} for both early times (on the unstable manifold of \PkD{1}) and late times (on the stable manifold of \PkD{1}).}
{Refining the estimate for $\eta_1$ using bisection revealed that ${\bf u}^-(\eta_1,t)$ instead converges to \EkD{3} (with $D_0 = 5\times10^{-4}$), approximating a heteroclinic connection \DCn{16} (blue curve in \reffig{homoclinic}). 

We found that trajectories ${\bf u}^-({\eta,t})$ in the interval marked $(\eta_3,\eta_4)$ in \reffig{s_vs_eta} also converge to \EkD{0}, after a brief turbulent excursion. These trajectories constitute a one-parameter family of connections \DCn{17} from \PkD{1} to \EkD{0}. Using bisection, we identified that ${\bf u}^-(\eta_3,t)$ is another homoclinic orbit (\DCn{18}) of \PkD{1} while ${\bf u}^-(\eta_4,t)$ is a heteroclinic connection (\DCn{19}) from \PkD{1} to \EkD{3}. The shapes of  \DCn{17}, \DCn{18}, and \DCn{19} are very similar to that of \DCn{14}, \DCn{16}, and \DCn{15}, respectively, and hence are not shown.
Note that we have so far inspected state space  along manifold trajectories to detect signatures of  dynamical connections to EQs. An alternative metric, which allows one to identify signatures of close passes to both EQs and POs is discussed in more detail in  Appendix \ref{sec:recurrence}. Trajectories ${\bf u}^-({\eta,t})$ originating at \PkD{1} are analyzed using both metrics and the results are compared in \reffig{rec}.

\subsection{Connections originating at \PkD{2}}\label{sec:po1b}

\begin{figure}[!tp]
\centering
{\includegraphics[width=3.3in,valign=c]{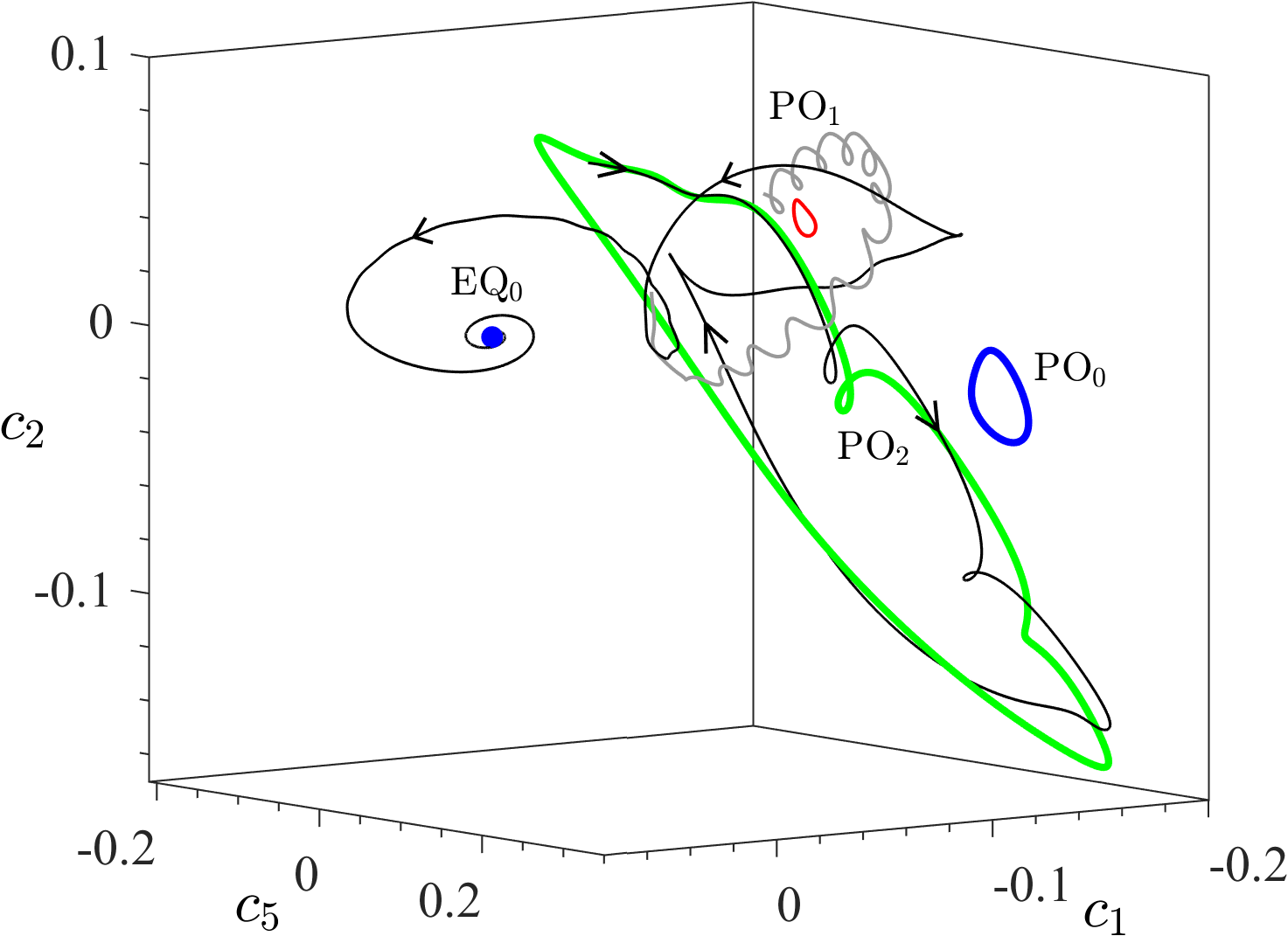}}
\caption{\label{fig:PO1B_to_E0nPO1A} Heteroclinic connections  from \PkD{2} (green loop) to \EkD{0} (\DCn{20}, black curve) and \PkD{1} (\DCn{21}, gray curve). {\DCn{21} eventually converges to \PkD{1} tracing out a helical path similar to that} shown in \reffig{projection_bisection}.} 
\end{figure}

Visual inspection of {state space} speed and recurrence plots  (e.g., \reffig{rec}) revealed that some turbulent trajectories ${\bf u}^-(\eta,t)$ from \PkD{1} shadow an unstable periodic orbit \PkD{2}. 
{This orbit, which we computed using a Newton solver, has a period $T = 4.61\tau_c$ and is only moderately repelling in $\mathcal{S}$; its leading Floquet exponents are $\lambda_1 = 0.044$, $\lambda_{2,3} =  -0.019 \pm 0.016i$.
However,  we did not find evidence of a short heteroclinic connection from \PkD{1} to \PkD{2}, i.e.,  ${\bf u}^-(\eta,t)$ for $t/\tau_c\in(15,35)$ did not approach \PkD{2} very closely.
Hence, we  tested {whether} a connection instead exists from \PkD{2} to \PkD{1}. 

The 2D unstable manifold of \PkD{2} is composed of two sets of trajectories ${\bf u}^\pm(\xi,t)$ that start from initial conditions ${\bf u}^\pm(\xi,0)$ constructed using \refeqs{1dm_upo}. Here,  $\xi$ (instead of $\eta$) parametrizes states along \PkD{2} as well as the trajectories ${\bf u}^\pm(\xi,t)$. 
Unlike {the unstable manifold of} \PkD{1}, we found that trajectories in both ${\bf u}^+(\xi,t)$ and ${\bf u}^-(\xi,t)$ display turbulent evolution.
Hence, we analyzed  ${\bf u}^-(\xi,t)$ as well as ${\bf u}^+(\xi,t)$ for signatures of dynamical connections; the plot of $s_m(\xi)$ for each set is included as Fig.\,S1 in the supplementary material. 
Inspecting {state space speed (as well as recurrence) plots for each trajectory and following the procedure outlined in the previous sections,} we identified a family of heteroclinic connections from \PkD{2} to {\EkD{0} (\DCn{20}) and two isolated connections to \PkD{1} (\DCn{21}, \DCn{22}) that lie at the boundary of \DCn{20}. A connection from the family \DCn{20} (black curve) and the isolated connection \DCn{21} (gray curve) are shown in \reffig{PO1B_to_E0nPO1A}.} 
The projection coordinates {here are the same as in \reffig{projection_bisection} and the viewing angle is similar.}  
We note that \DCn{20}, \DCn{21}, and \DCn{22} (not shown) approach their respective destinations {quickly} after leaving the neighborhood of \PkD{2}.

\begin{figure}[!tp]
\includegraphics[width=3.3in]{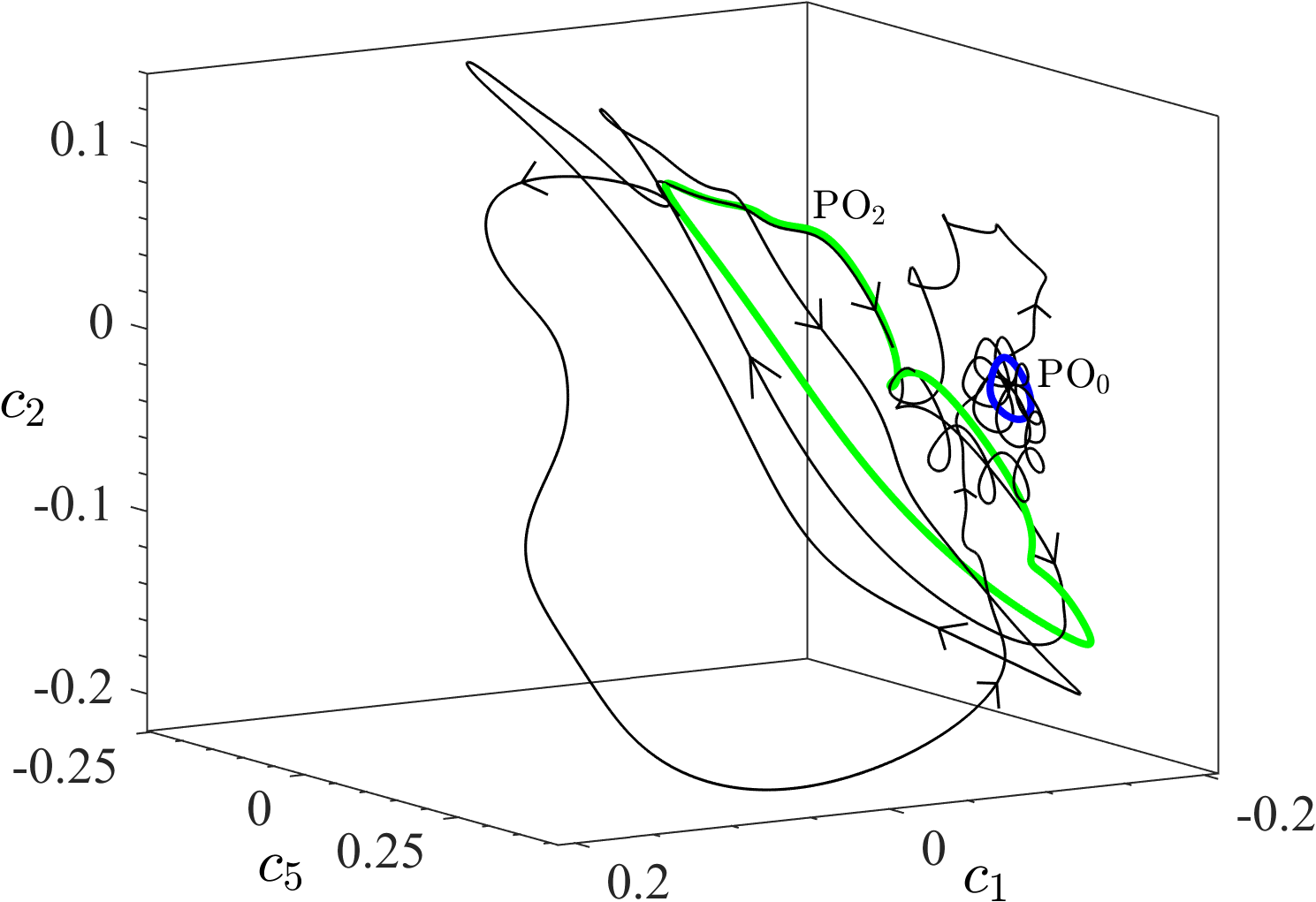}
\begin{picture}(0,0)
\put(-210,42){\fbox{\includegraphics[width=1.75cm]{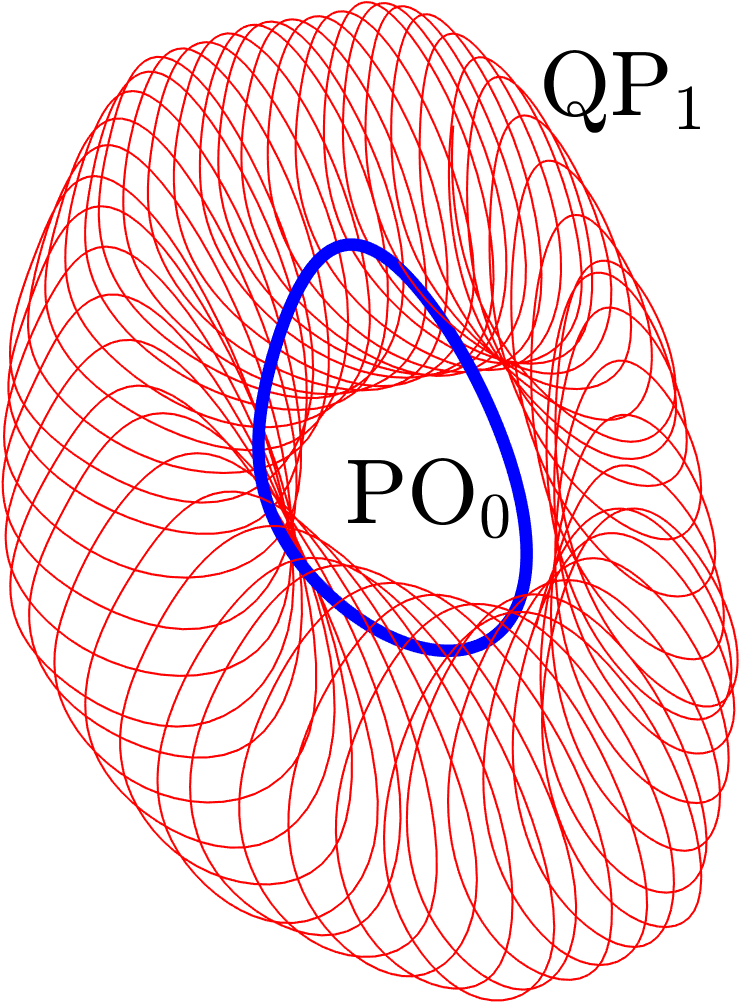}}}
\end{picture}
\caption{\label{fig:PO2_to_PO0} Heteroclinic connection \DCn{23} from \PkD{2} (green loop)  to \PkD{0} (blue loop). {The inset shows the unstable quasi-periodic orbit \TkD{1} (red curve) that lies near stable \PkD{0}.}} 
\end{figure}

We also found that a narrow band of trajectories ${\bf u}^+(\xi,t)$ for $\xi \in (\xi_1,\xi_2)$ from \PkD{2} approach a periodic orbit \PkD{0} shown in  \reffig{PO1B_to_E0nPO1A}. 
\PkD{0} has a period $T = 1.22\tau_c$ and is stable in $\mathcal{S}$, with its leading Floquet exponent being $\lambda_{1,2} =  -0.0026 \pm 0.059i$. 
Hence, a one-parameter family of connections \DCn{23} from \PkD{2} to \PkD{0} was computed by evolving ${\bf u}^+(\xi,t)$ for $\xi \in (\xi_1,\xi_2)$ for $t\approx 400\tau_c$.  The state space projection of a trajectory from the family \DCn{23} is shown in \reffig{PO2_to_PO0}. Since \PkD{0} is stable in $\mathcal{S}$,  \DCn{23} is a 2D {manifold} and trajectories that correspond to the left ($\xi_1$) and right ($\xi_2$) edges separate trajectories converging to \PkD{0} from those which become turbulent.
Using bisection, we found that both ${\bf u}^+(\xi_1,t)$, ${\bf u}^+(\xi_2,t)$ approach {a 2-torus representing an unstable quasi-periodic orbit (QPO)} \TkD{1} shown in \reffig{PO2_to_PO0} (inset). Hence,  ${\bf u}^+(\xi_1,t)$, ${\bf u}^+(\xi_2,t)$  correspond to connections \DCn{24}, \DCn{25} from \PkD{2} to \TkD{1}.}

The state space speed for ${\bf u}^+(\xi_1,t)$ clearly shows {dynamics with two different frequencies} over a time interval $70\tau_c$ (Fig.\,S2 in supplementary material).
Computing recurrence plots for this segment of ${\bf u}^+(\xi_1,t)$, we estimated that the periods associated with large and small loops of \TkD{1} are $T_1/\tau_c = 12.21\pm 0.018 $ and $T_2/\tau_c = 1.21\pm 0.05$, respectively. Since {the ratio} $T_1/T_2 \approx 10.05$ is close to an integer, we also used the Newton-Krylov solver to test whether \TkD{1} is instead a periodic orbit with period $T\approx T_1,\, 2T_1,$ or $3T_1$; in all cases the solver failed to converge, {suggesting that \TkD{1} is indeed a QPO.} 
The procedure we used to compute \TkD{1} also suggests that this solution (just like \PkD{1}) {possesses just one unstable direction in $\mathcal S$ and its stable manifold separates initial conditions that quickly approach \PkD{0} from those exhibiting   transient turbulence.}
Computing the stability exponents of \TkD{1}, however, is beyond the scope of this study. 
Lastly, the trajectories from \TkD{1} that uneventfully converge to \PkD{0} constitute a two-parameter family of connections forming a 3D manifold (\DCn{26}). 

\subsection{Structural Stability of Dynamical Connections}\label{sec:lim_cycle}

All dynamical connections we reported so far were computed at $Re = 22.05$. 
One may ask if these connections are robust to small changes in $Re$, i.e., {whether these} connections are structurally stable \cite{smale_1967}. 
{Recall that a connection from ECS$_{-\infty}$ to ECS$_{\infty}$  is formed by an intersection of the unstable manifold of ECS$_{-\infty}$ and the stable manifold of ECS$_\infty$, with respective dimensions $d_u^{-\infty}$ and $d_s^{\infty}$. If these two manifolds intersect (in our case we have shown that they do), their intersection will generically be of dimension $d=d_u^{-\infty}+d_s^{\infty}-N$, where $N$ is the dimension of the state space (in our case $\mathcal{S}$). We should have $d>0$ for the connection to be structurally stable.

Since both $d_s^{\infty}$ and $N$ are typically very large (or infinite), $d$ can be expressed in a more convenient form in terms of the co-dimension  $k_s^\infty=N-d_s^\infty$ of the stable manifold of ECS$_\infty$ \cite{halcrow_2009}.
If $N_u^\infty$, $N_s^\infty$, and $N_m^\infty$ are the number of unstable, stable, and marginal directions of ECS$_\infty$,  then  $N_u^\infty+N_s^\infty+N_m^\infty = N$ and $d_s^\infty = N_s^\infty + N_m^\infty$. Hence, $k_s^\infty = N_u^\infty$ which yields $d = d_u^{-\infty}-N_u^\infty$. The criterion  ($d>0$) for the structural stability of a connection is then simply $d_u^{-\infty} > N_u^\infty$. 
\reftable{connection_summary} lists the various connections we computed and the corresponding dimensions for each connection. The entire network of connections is also shown in schematic form in \reffig{topology}. 
Clearly, all of our connections satisfy the structural stability criterion. 
We also numerically validated the structural stability of the connections computed at $Re = 22.05$ by continuing the ECSs for $Re \in [21.85,22.2]$ and analyzing their unstable manifolds.  
For $Re\in [21.85,22.2]$ the number of unstable directions of {all} the ECSs we computed (except \PkD{2}) remained unchanged. \PkD{2} exists only for $Re \geq 22.0$ and its unstable manifold was  analyzed only at $Re = 22.2$.

\begin{figure}[!tp]
\centering
\fbox{\includegraphics[width=3.3in,valign=c]{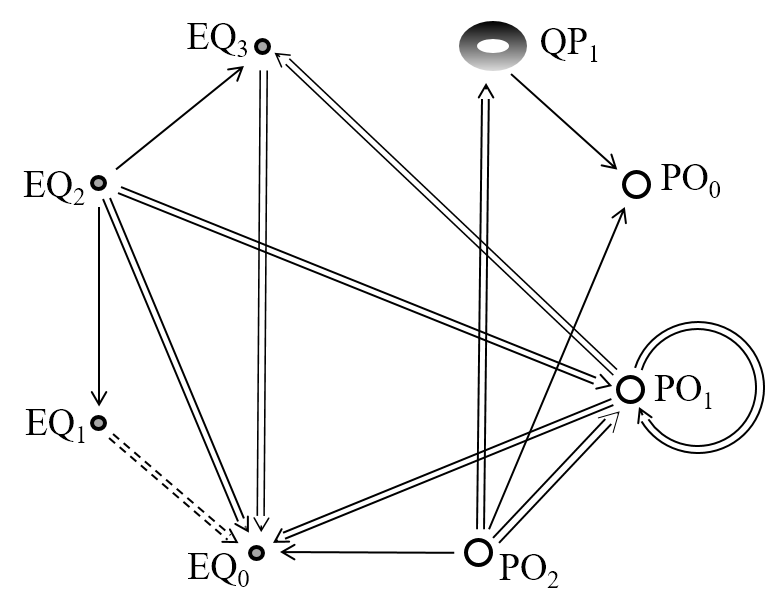}}
\caption{\label{fig:topology} Topology of connections. Double lines represent multiple distinct connections. Solid (dashed) lines represent short (long) connections.}
\end{figure}
\setlength{\tabcolsep}{5pt}
\begin{table}[!ht]
\centering
\ra{1.3}
\begin{tabular}{|c|c|c|c|c|c|}	
\hline
{} & ECS$_{-\infty}$ & ECS$_\infty$ & d & $d_u^{-\infty}$ & $k_s^\infty$  \\
\midrule
\hline
\DCn{1}, \DCn{2} & \EkD{2} & \EkD{0} & 2 & 2 & 0 \\
\DCn{3} & \EkD{2} & \EkD{1} & 1 & 2 & 1 \\
\DCn{4}, \DCn{5} & \EkD{1} & \EkD{0} & 1 & 1 & 0 \\
\DCn{6}--\DCn{9}  & \EkD{2} & \PkD{1} & 1 & 2 & 1 \\
\DCn{10} & \EkD{2} & \EkD{3} & 1 & 2 & 1 \\
\DCn{11}, \DCn{12} & \EkD{3} & \EkD{0} & 1 & 1 & 0 \\
{\DCn{13}, \DCn{14}, \DCn{17}} & \PkD{1} & \EkD{0} & 2 & 2 & 0 \\
{\DCn{15}, \DCn{18}}  & \PkD{1} & \PkD{1} & 1 & 2 & 1 \\
{\DCn{16}, \DCn{19}} & \PkD{1} & \EkD{3} & 1 & 2 & 1 \\
\DCn{20} & \PkD{2} & \EkD{0} & 2 & 2 & 0 \\
\DCn{21}, \DCn{22} & \PkD{2} & \PkD{1} & 1 & 2 & 1 \\
\DCn{23} & \PkD{2} & \PkD{0} & 2 & 2 & 0 \\
\DCn{24}, \DCn{25} & \PkD{2} & \TkD{1} & 1 & 2 & 1 \\
\DCn{26} & \TkD{1} & \PkD{0} & 3 & 3 & 0 \\
\hline
\bottomrule
\end{tabular}
\caption{\label{table:connection_summary} Dynamical connections computed at $Re = 22.05$. }
\end{table}

\subsection{Transient Turbulence}\label{sec:transient}

\begin{figure*}[!tp]
\centering
\includegraphics[width=7in]{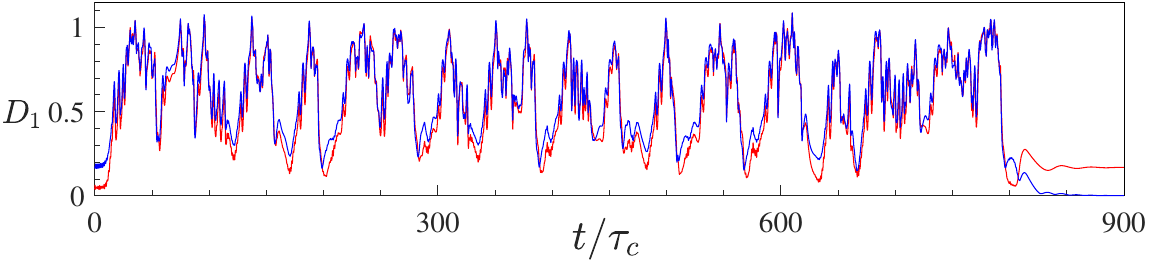}
\caption{\label{fig:transient}  Instantaneous distances from  a long transient turbulent  trajectory to \PkD{1} ($D_1$, red) and  \EkD{0} ($D_0$, blue). Repeated close passes to \PkD{1} suggest that the homoclinic tangle between the stable and unstable manifolds of \PkD{1} underpins transient turbulence in symmetric subspace. \authedits{Vanishing of $D_0$ implies convergence to \EkD{0}}.} 
\end{figure*}

{As we discussed previously, turbulence in the symmetry subspace $\mathcal S$ is transient, with most initial conditions eventually ``relaminarizing'' by converging to the stable equilibrium \EkD{0}.
While some states relaminarize relatively quickly,  others stay turbulent for a significant interval of time.
It is natural to ask what geometric structures are responsible for maintaining turbulent flow and for relaminarization.
Our results show that the periodic orbit \PkD{1} plays a key role in both processes.
As we have demonstrated, \PkD{1} possesses (at least) two distinct homoclinic connections (\DCn{15}, \DCn{18}).
The presence of homoclinic connections, which lie at the intersection of the stable and unstable manifold of \PkD{1}, suggests that these manifolds intersect and form a homoclinic tangle.
The tangle explains the fractal nature of the minimal state space speed shown in \reffig{s_vs_eta} which results from stretching and folding of the unstable manifold.
More importantly, it implies the presence of a chaotic set (a chaotic repeller in our case) anchored by \PkD{1} as well as the presence of arbitrarily long periodic orbits that visit the neighborhood of \PkD{1} \cite{smale_1967}.
These are precisely the ingredients required for transient turbulence.}

{To show that it is indeed the homoclinic tangle associated with \PkD{1} that underlies transient turbulence in our system, we computed the distance $D_1$ ($D_0$) from  \PkD{1} (\EkD{0}) to  a particular long turbulent trajectory.
As \reffig{transient} demonstrates, the trajectory returns to the vicinity of \PkD{1} {many} times before finally relaminarizing i.e., converging to \EkD{0} ($D_0 \rightarrow 0$).
Furthermore, just before relaminarization, this trajectory comes very close to \PkD{1}, which suggests \PkD{1} plays an important role in this process. 
It should be pointed out that not all trajectories approach \PkD{1} closely just before relaminarization. 
For instance, some trajectories pass through the neighborhood of \EkD{3} instead, as  \reffig{homoclinic} illustrates.
Both \PkD{1} and \EkD{3} have stable manifolds with co-dimension one; states on one side of these manifolds relaminarize almost immediately and those on the other side exhibit transient turbulence. 
Hence, these two stable manifolds form {portions of} a local boundary between ``laminar"  and  ``turbulent" {flows}. 
This analogy, however, is not perfect since the chaotic set {underlying the turbulent transient} is not an attractor.

\authedits{It should be mentioned that non-uniqueness of edge states lying on a ``laminar-turbulent'' boundary has been previously reported for other turbulent flows as well. For instance, Kerswell \cite{kerswell_2007} and Duguet \etal \cite{duguet_2008a}  have identified several edge states corresponding to different traveling wave solutions in a short periodic pipe. Our results provide further evidence that not only can multiple edge states coexist, they can be of different types (e.g., an EQ and a PO, in our case).}

The relationship between transient turbulence and chaotic repellers has been suggested previously, mainly based on indirect evidence -- the power law decay of the relaminarization times characterizing a memoryless process \cite{hof_2006,eckhardt2007dynamical,schneider2010transient,borrero_2010}.
Direct evidence, such as the presence of homoclinic tangles \cite{vanveen_2011b, budanur_2019} or a period-doubling cascade \cite{kreilos_2012}, is  more recent.
Moreover, while the dynamics and stability in systems with heteroclinic cycles have been explored previously \cite{kirk_1994,krupa_1995}, there is currently very little understanding of dynamics in the presence of both heteroclinic and homoclinic connections.
Due to the relative simplicity of the numerical model and ease of experimental access, the system considered here is particularly attractive for studying the relation between transient dynamics, relaminarization, and the structure of connections.

\section{Summary and Conclusions}\label{sec:summary}

Several recent studies on a dynamical description of fluid turbulence focused on ECSs and how they shape the state space geometry in their neighborhoods. 
However, complete understanding of turbulence requires a global picture which explains how the flow moves between neighborhoods of ECSs.
Such a picture can be considered as a coarse description of the dynamics, in the spirit of symbolic dynamics, analogous to a 
route network where ECSs serve as nodes and dynamical (homo/heteroclinic) connections as links connecting the nodes. This study describes the first rigorous attempt to construct such a network for a turbulent fluid flow. We identified eight nodes and several tens of connections between them, far more than any study to date.
Moreover, while most previous studies computed connections between ECSs of the same type (primarily EQs), we have identified connections between three different types of ECSs: EQ, PO, and QPO. Indeed, this is the first study to compute connections involving QPOs. 
We have also demonstrated, for the first time, the existence of higher (two/three) dimensional connections between ECSs, i.e., continua of trajectories from one ECS to another.

Despite the limited attention they have received, dynamical connections {can} play a very important role in turbulent evolution \cite{gibson_2008}. 
For instance, dynamically dominant ECSs in the Kolmogorov-like flow  are equilibria \cite{suri_2018}. Being fixed points in the state space, EQs cannot guide turbulent trajectories in their neighborhoods in the same way POs or QPOs do. Therefore,  
connections between EQs (as well as other types of ECSs) 
become the dynamically dominant solutions that guide turbulent trajectories, shaping their evolution locally.
Even in systems where the dominant solutions are POs or QPOs,
ECSs constrain the dynamics only locally in state space and over short intervals of time. The network of connections, on the other hand, constrains the dynamics globally in state space and over arbitrarily long time intervals.

Identifying the connection network has potential applications such as forecasting \cite{suri_2017a} and control of turbulent flows \cite{hof_2010,kuhnen_2018}. Even though quantitative prediction has a time horizon set by the leading Lyapunov exponents that characterize the sensitivity to initial conditions, qualitative predictions do not have this limitation. 
In principle, prediction of extreme events can be made  based on the connectivity of different ECSs.
Identifying the connection network can also facilitate ``low-energy'' control of turbulent flows, 
where small perturbations to the flow result in its subsequent (natural) evolution towards a particular ECS or region of state space with desired behavior \cite{pringle_2012}.
Connections can also provide new insight into laminar-turbulent transition in wall-bounded 3D shear flows \cite{budanur_2017b,budanur_2019}.

Lastly, we point out that constraining the dynamics to a symmetry invariant subspace lowered the dimensionality of  unstable manifolds and dramatically simplified the procedure of computing connections between ECSs.
Whether these connections are dynamically relevant in the full state space requires further exploration.
Currently, we are lacking robust numerical methods for computing connections between ECSs with more than one or two unstable directions. 
Some approaches, such as adjoint looping have shown promise \cite{farano_2018a}, but whether they present a viable option for computing connections between different types of ECS remains an open problem. 

\begin{acknowledgments}
Suri thanks  Burak Budanur for useful discussions and Bj\"{o}rn Hof for providing financial support over the duration of this study.  
Suri acknowledges funding from the European Union’s Horizon 2020 research and innovation program under the Marie Sk\l{}odowska-Curie grant agreement No 754411. MS and RG acknowledge funding from the National Science Foundation (CMMI-1234436, DMS-1125302, CMMI-1725587) and Defense Advanced Research Projects Agency (HR0011-16-2-0033).
\end{acknowledgments}

\appendix
\section{State Space Projections}\label{sec:projections}
\authedits{
In this appendix we briefly describe the procedure used to project the state space onto a low-dimensional subspace spanned by the eigenvectors {(Floquet vectors)}  $\uv{e}_i$ of an equilibrium {(periodic orbit)}. Each trajectory ${\bf u}(t)$ is expressed as a linear combination of the vectors ${\uv{e}_i}$ as follows:
\begin{equation}\label{eq:lincomb}
{\bf u}(t) = {\bf u}_{ecs}+ \sum\limits_ia_i(t)\uv{e}_i.
\end{equation}
{Here, ${\bf u}_{ecs}$ corresponds to an equilibrium  (e.g., \EkD{0}) or a point on a periodic orbit (e.g., \PkD{1})}. 
The coefficients $a_i(t)$ are computed using the scalar product
\begin{equation}
a_{i}(t) = \uv{e}_i^\dagger\cdot({\bf u}(t)-{\bf u}_{ecs}), 
\end{equation}
where $\uv{e}_i^\dagger$ is the adjoint eigenvector {(Floquet vector)} such that  $\uv{e}_i^\dagger\cdot\uv{e}_j = \delta_{ij}$ (Kronecker delta). Typically, the vectors $\uv{e}_i$ are not  orthonormal ($\uv{e}_i\cdot\uv{e}_j \neq \delta_{ij}$). Hence, we construct orthonormalized vectors $\uv{e}_j'$  such that  
\begin{equation}
\uv{e}_i =  \sum\limits_jT_{ij}\uv{e}_j',
\end{equation}
where, the matrix elements $T_{ij} = \uv{e}_i\cdot\uv{e}_j'$ can be computed using the orthonormality condition $\uv{e}_i'\cdot\uv{e}_j' = \delta_{ij}$. 
The normalized components 
\begin{equation}
c_j = \sum\limits_ia_iT_{ij}/D_c
\end{equation}
along vectors $\uv{e}_i'$ are plotted to generate state space projections. 
Here, $D_c$ is the empirically estimated largest separation between two states {on} a long turbulent trajectory (cf. \refsec{2dm_e2}).
}

\section{Recurrence analysis}\label{sec:recurrence}

\begin{figure*}[!htbp]
\subfloat[]{\includegraphics[width=3.4in,valign=c]{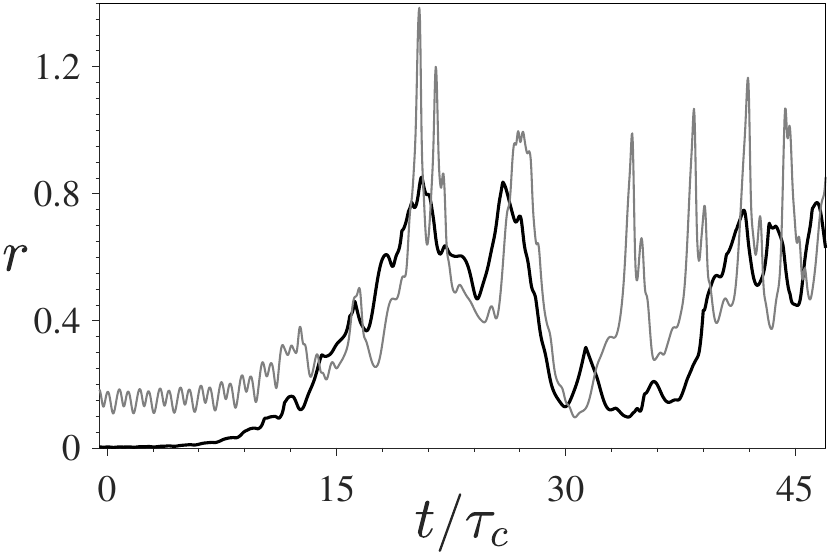}} \hspace{1mm} 
\subfloat[]{\includegraphics[width=3.4in,valign =c]{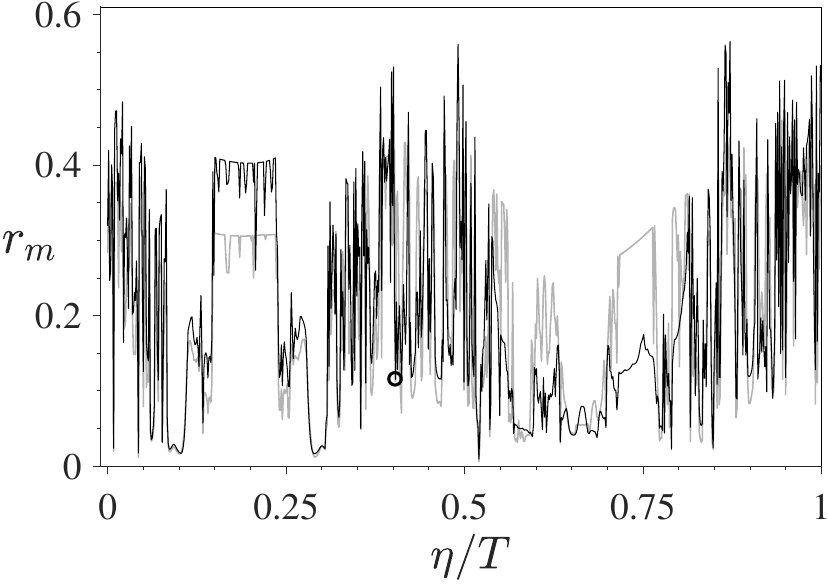}}
\caption{\label{fig:rec} Recurrence analysis to detect signatures of both EQs and POs (a) Recurrence  (black curve) and speed (gray curve) plots for a trajectory that originates at \PkD{1} and visits the neighborhoods of \EkD{0} ($t/\tau_c = 30$) and \PkD{2} ($t/\tau_c\in[33,40]$). (b) Minimum recurrence $r_m$ (black curve) and minimum speed $s_m$ (gray curve) for each trajectory ${\bf u}^-({\eta,t})$ from \PkD{1}. Trajectories with $r_m\ll1$ ($s_m\ll 1$) correspond to dynamical connections from \PkD{1} to EQs/POs (EQs). $s_m$ was scaled such that $r_m$, $s_m$ have the same mean. Black circle indicates $r_m$ for the trajectory shown in panel (a).} 
\end{figure*}

In sections \ref{sec:2dm_e2}-\ref{sec:po1b} we identified close passes to EQs by computing state space speed $s(t)$ along turbulent trajectories (cf. \reffig{sss}). 
However, $s(t)$ is not zero (or constant) for POs, {so detecting close passes to POs requires visual inspection of} speed plots {to find intervals of} oscillatory behavior.
To  address this shortcoming, we tested recurrence analysis \cite{eckmann_1987} in a form similar to that discussed in Duguet \etal \cite{duguet_2008b}. For each trajectory ${\bf u}(t)$ we computed the {normalized recurrence function}
\begin{equation}\label{eq:rec}
r(t) = \min\limits_{\tau\in[\tau_l,\tau_h]} \frac{\|{\bf u}(t)-{\bf u}(t+\tau)\|}{D_c},
\end{equation}
where $0<\tau_l<\tau_h$ are appropriately chosen constants.
Low values of $r$ indicate that the flow field  at an instant $t$ nearly recurs during a later interval $[t+\tau_l, t+\tau_h]$. 
Unlike state space speed, $r(t)=0$ for {${\bf u}(t)$ representing} \textit{both} EQs and POs. {In the former case $\tau_l$ and $\tau_h$ are arbitrary, while in the latter case the period $T$ of the orbit should lie inside the interval $[\tau_l,\tau_h]$.}

Since a turbulent trajectory {shadowing} an ECS mimics its spatiotemporal behavior, {intervals where $r(t)\ll 1$ correspond to the trajectory visiting neighborhoods of  EQs or POs.} To identify such visits, however, we should restrict how near (or far) into the future we search for recurrence.  Choosing {$\tau_l$ far smaller than the correlation time $\tau_c$} leads to spurious self-recurrence since ${\bf u}(t)$ and ${\bf u}(t+\tau)$ do not differ appreciably; the extreme case being $r(t) = 0$ when $\tau_l = 0$. 
The upper bound $\tau_h$ restricts the search to short periodic orbits (which tend to be dynamically relevant \cite{cvitanovic_2010}) with period $T<\tau_h$. Besides, it also limits the overhead associated with computing $r(t)$. In our analysis we chose $\tau_l = \tau_c$ and $\tau_h = 5\tau_c$.

\reffig{rec}(a) shows recurrence (black curve) and speed (gray curve) plots for a turbulent  trajectory ${\bf u}(t)$ that originates at \PkD{1}.
Initially,  ${\bf u}(t)$ {shadows} \PkD{1} and consequently $s(t)$ displays steady oscillations {about a finite value}; in contrast,  $r(t)$ {almost vanishes}.  
After a brief turbulent excursion characterized by $r$ increasing to $O(1)$, ${\bf u}(t)$ visits the neighborhoods of \EkD{0} at $t/\tau_c\approx 30$ and \PkD{2} for $33 \lesssim t/\tau_c \lesssim 40$.
Both these close passes correspond to {$r$ decreasing to well below unity}. 
Hence, to identify signatures of dynamical connections from \PkD{1} to both EQs and POs, we computed $r(\eta,t)$ for each  ${\bf u}^-(\eta,t)$ in the unstable manifold of \PkD{1} (cf. \refsec{separatrix}). 
We then computed $r_m(\eta) = \min_t r(\eta,t)$ for $t>15\tau_c$, i.e., after each trajectory initially leaves the neighborhood of \PkD{1}. 
{The results are shown in \reffig{rec}(b) which compares $r_m(\eta)$ (black curve) with $s_m(\eta)$ (gray curve) for each trajectory ${\bf u}^-(\eta,t)$. 
Clearly, the prominent minima of the two metrics align,  which suggests that recurrence-based analysis is capable of successfully identifying signatures of close passes to both EQs and POs. However,  it is a slightly more expensive method to identify connections, compared with the minimal state space speed.}


\begin{thebibliography}{65}%
\makeatletter
\providecommand \@ifxundefined [1]{%
\@ifx{#1\undefined}
}%
\providecommand \@ifnum [1]{%
\ifnum #1\expandafter \@firstoftwo
\else \expandafter \@secondoftwo
\fi
}%
\providecommand \@ifx [1]{%
\ifx #1\expandafter \@firstoftwo
\else \expandafter \@secondoftwo
\fi
}%
\providecommand \natexlab [1]{#1}%
\providecommand \enquote  [1]{``#1''}%
\providecommand \bibnamefont  [1]{#1}%
\providecommand \bibfnamefont [1]{#1}%
\providecommand \citenamefont [1]{#1}%
\providecommand \href@noop [0]{\@secondoftwo}%
\providecommand \href [0]{\begingroup \@sanitize@url \@href}%
\providecommand \@href[1]{\@@startlink{#1}\@@href}%
\providecommand \@@href[1]{\endgroup#1\@@endlink}%
\providecommand \@sanitize@url [0]{\catcode `\\12\catcode `\$12\catcode
`\&12\catcode `\#12\catcode `\^12\catcode `\_12\catcode `\%12\relax}%
\providecommand \@@startlink[1]{}%
\providecommand \@@endlink[0]{}%
\providecommand \url  [0]{\begingroup\@sanitize@url \@url }%
\providecommand \@url [1]{\endgroup\@href {#1}{\urlprefix }}%
\providecommand \urlprefix  [0]{URL }%
\providecommand \Eprint [0]{\href }%
\providecommand \doibase [0]{http://dx.doi.org/}%
\providecommand \selectlanguage [0]{\@gobble}%
\providecommand \bibinfo  [0]{\@secondoftwo}%
\providecommand \bibfield  [0]{\@secondoftwo}%
\providecommand \translation [1]{[#1]}%
\providecommand \BibitemOpen [0]{}%
\providecommand \bibitemStop [0]{}%
\providecommand \bibitemNoStop [0]{.\EOS\space}%
\providecommand \EOS [0]{\spacefactor3000\relax}%
\providecommand \BibitemShut  [1]{\csname bibitem#1\endcsname}%
\let\auto@bib@innerbib\@empty
\bibitem [{\citenamefont {Kerswell}(2005)}]{kerswell_2005}%
\BibitemOpen
\bibfield  {author} {\bibinfo {author} {\bibfnamefont {R.~R.}\ \bibnamefont
{Kerswell}},\ }\href@noop {} {\bibfield  {journal} {\bibinfo  {journal}
{Nonlinearity}\ }\textbf {\bibinfo {volume} {18}},\ \bibinfo {pages} {R17}
(\bibinfo {year} {2005})}\BibitemShut {NoStop}%
\bibitem [{\citenamefont {Kawahara}\ \emph {et~al.}(2012)\citenamefont
{Kawahara}, \citenamefont {Uhlmann},\ and\ \citenamefont {van
Veen}}]{kawahara_2012}%
\BibitemOpen
\bibfield  {author} {\bibinfo {author} {\bibfnamefont {G.}~\bibnamefont
{Kawahara}}, \bibinfo {author} {\bibfnamefont {M.}~\bibnamefont {Uhlmann}}, \
and\ \bibinfo {author} {\bibfnamefont {L.}~\bibnamefont {van Veen}},\
}\href@noop {} {\bibfield  {journal} {\bibinfo  {journal} {Annu. Rev. Fluid
Mech.}\ }\textbf {\bibinfo {volume} {44}},\ \bibinfo {pages} {203} (\bibinfo
{year} {2012})}\BibitemShut {NoStop}%
\bibitem [{\citenamefont {Hof}\ \emph {et~al.}(2004)\citenamefont {Hof},
\citenamefont {van Doorne}, \citenamefont {Westerweel}, \citenamefont
{Nieuwstadt}, \citenamefont {Faisst}, \citenamefont {Eckhardt}, \citenamefont
{Wedin}, \citenamefont {Kerswell},\ and\ \citenamefont {Waleffe}}]{hof_2004}%
\BibitemOpen
\bibfield  {author} {\bibinfo {author} {\bibfnamefont {B.}~\bibnamefont
{Hof}}, \bibinfo {author} {\bibfnamefont {C.~W.~H.}\ \bibnamefont {van
Doorne}}, \bibinfo {author} {\bibfnamefont {J.}~\bibnamefont {Westerweel}},
\bibinfo {author} {\bibfnamefont {F.~T.~M.}\ \bibnamefont {Nieuwstadt}},
\bibinfo {author} {\bibfnamefont {H.}~\bibnamefont {Faisst}}, \bibinfo
{author} {\bibfnamefont {B.}~\bibnamefont {Eckhardt}}, \bibinfo {author}
{\bibfnamefont {H.}~\bibnamefont {Wedin}}, \bibinfo {author} {\bibfnamefont
{R.~R.}\ \bibnamefont {Kerswell}}, \ and\ \bibinfo {author} {\bibfnamefont
{F.}~\bibnamefont {Waleffe}},\ }\href@noop {} {\bibfield  {journal} {\bibinfo
{journal} {Science}\ }\textbf {\bibinfo {volume} {305}},\ \bibinfo {pages}
{1594} (\bibinfo {year} {2004})}\BibitemShut {NoStop}%
\bibitem [{\citenamefont {de~Lozar}\ \emph {et~al.}(2012)\citenamefont
{de~Lozar}, \citenamefont {Mellibovsky}, \citenamefont {Avila},\ and\
\citenamefont {Hof}}]{lozar_2012}%
\BibitemOpen
\bibfield  {author} {\bibinfo {author} {\bibfnamefont {A.}~\bibnamefont
{de~Lozar}}, \bibinfo {author} {\bibfnamefont {F.}~\bibnamefont
{Mellibovsky}}, \bibinfo {author} {\bibfnamefont {M.}~\bibnamefont {Avila}},
\ and\ \bibinfo {author} {\bibfnamefont {B.}~\bibnamefont {Hof}},\
}\href@noop {} {\bibfield  {journal} {\bibinfo  {journal} {Phys. Rev. Lett.}\
}\textbf {\bibinfo {volume} {108}},\ \bibinfo {pages} {214502} (\bibinfo
{year} {2012})}\BibitemShut {NoStop}%
\bibitem [{\citenamefont {Suri}\ \emph {et~al.}(2017)\citenamefont {Suri},
\citenamefont {Tithof}, \citenamefont {Grigoriev},\ and\ \citenamefont
{Schatz}}]{suri_2017a}%
\BibitemOpen
\bibfield  {author} {\bibinfo {author} {\bibfnamefont {B.}~\bibnamefont
{Suri}}, \bibinfo {author} {\bibfnamefont {J.}~\bibnamefont {Tithof}},
\bibinfo {author} {\bibfnamefont {R.~O.}\ \bibnamefont {Grigoriev}}, \ and\
\bibinfo {author} {\bibfnamefont {M.~F.}\ \bibnamefont {Schatz}},\ }\href
{\doibase 10.1103/PhysRevLett.118.114501} {\bibfield  {journal} {\bibinfo
{journal} {Phys. Rev. Lett.}\ }\textbf {\bibinfo {volume} {118}},\ \bibinfo
{pages} {114501} (\bibinfo {year} {2017})}\BibitemShut {NoStop}%
\bibitem [{\citenamefont {Hopf}(1948)}]{hopf_1948}%
\BibitemOpen
\bibfield  {author} {\bibinfo {author} {\bibfnamefont {E.}~\bibnamefont
{Hopf}},\ }\href@noop {} {\bibfield  {journal} {\bibinfo  {journal} {Commun.
Pur. Appl. Math.}\ }\textbf {\bibinfo {volume} {1}},\ \bibinfo {pages} {303}
(\bibinfo {year} {1948})}\BibitemShut {NoStop}%
\bibitem [{\citenamefont {Gibson}\ \emph {et~al.}(2008)\citenamefont {Gibson},
\citenamefont {Halcrow},\ and\ \citenamefont {Cvitanovi{\'c}}}]{gibson_2008}%
\BibitemOpen
\bibfield  {author} {\bibinfo {author} {\bibfnamefont {J.~F.}\ \bibnamefont
{Gibson}}, \bibinfo {author} {\bibfnamefont {J.}~\bibnamefont {Halcrow}}, \
and\ \bibinfo {author} {\bibfnamefont {P.}~\bibnamefont {Cvitanovi{\'c}}},\
}\href@noop {} {\bibfield  {journal} {\bibinfo  {journal} {J. Fluid Mech.}\
}\textbf {\bibinfo {volume} {611}},\ \bibinfo {pages} {107} (\bibinfo {year}
{2008})}\BibitemShut {NoStop}%
\bibitem [{Note1()}]{Note1}%
\BibitemOpen
\bibinfo {note} {See supplementary material at  [xxx] for (a) Videos showcasing side-by-side evolution in state space and physical space for connections \DCn{1}, \DCn{3}, \DCn{6}, \DCn{10}, \DCn{11}, \DCn{13}, \DCn{15}, \DCn{20}, and \DCn{23} (b) State space speed analysis of trajectories in the unstable manifold of \PkD{2}.}\BibitemShut {Stop}%
\bibitem [{\citenamefont {Duguet}\ \emph
{et~al.}(2008{\natexlab{a}})\citenamefont {Duguet}, \citenamefont {Willis},\
and\ \citenamefont {Kerswell}}]{duguet_2008b}%
\BibitemOpen
\bibfield  {author} {\bibinfo {author} {\bibfnamefont {Y.}~\bibnamefont
{Duguet}}, \bibinfo {author} {\bibfnamefont {A.~P.}\ \bibnamefont {Willis}},
\ and\ \bibinfo {author} {\bibfnamefont {R.~R.}\ \bibnamefont {Kerswell}},\
}\href@noop {} {\bibfield  {journal} {\bibinfo  {journal} {J. Fluid Mech.}\
}\textbf {\bibinfo {volume} {613}},\ \bibinfo {pages} {255} (\bibinfo {year}
{2008}{\natexlab{a}})}\BibitemShut {NoStop}%
\bibitem [{\citenamefont {Viswanath}\ and\ \citenamefont
{Cvitanovic}(2009)}]{viswanath_2009}%
\BibitemOpen
\bibfield  {author} {\bibinfo {author} {\bibfnamefont {D.}~\bibnamefont
{Viswanath}}\ and\ \bibinfo {author} {\bibfnamefont {P.}~\bibnamefont
{Cvitanovic}},\ }\href {\doibase 10.1017/S0022112009006041} {\bibfield
{journal} {\bibinfo  {journal} {J. Fluid Mech.}\ }\textbf {\bibinfo {volume}
{627}},\ \bibinfo {pages} {215–233} (\bibinfo {year} {2009})}\BibitemShut
{NoStop}%
\bibitem [{\citenamefont {Suri}\ \emph {et~al.}(2018)\citenamefont {Suri},
\citenamefont {Tithof}, \citenamefont {Grigoriev},\ and\ \citenamefont
{Schatz}}]{suri_2018}%
\BibitemOpen
\bibfield  {author} {\bibinfo {author} {\bibfnamefont {B.}~\bibnamefont
{Suri}}, \bibinfo {author} {\bibfnamefont {J.}~\bibnamefont {Tithof}},
\bibinfo {author} {\bibfnamefont {R.~O.}\ \bibnamefont {Grigoriev}}, \ and\
\bibinfo {author} {\bibfnamefont {M.~F.}\ \bibnamefont {Schatz}},\ }\href
{\doibase 10.1103/PhysRevE.98.023105} {\bibfield  {journal} {\bibinfo
{journal} {Phys. Rev. E}\ }\textbf {\bibinfo {volume} {98}},\ \bibinfo
{pages} {023105} (\bibinfo {year} {2018})}\BibitemShut {NoStop}%
\bibitem [{\citenamefont {Halcrow}\ \emph {et~al.}(2009)\citenamefont
{Halcrow}, \citenamefont {Gibson}, \citenamefont {Cvitanovi{\'c}},\ and\
\citenamefont {Viswanath}}]{halcrow_2009}%
\BibitemOpen
\bibfield  {author} {\bibinfo {author} {\bibfnamefont {J.}~\bibnamefont
{Halcrow}}, \bibinfo {author} {\bibfnamefont {J.~F.}\ \bibnamefont {Gibson}},
\bibinfo {author} {\bibfnamefont {P.}~\bibnamefont {Cvitanovi{\'c}}}, \ and\
\bibinfo {author} {\bibfnamefont {D.}~\bibnamefont {Viswanath}},\ }\href@noop
{} {\bibfield  {journal} {\bibinfo  {journal} {J. Fluid Mech.}\ }\textbf
{\bibinfo {volume} {621}},\ \bibinfo {pages} {365} (\bibinfo {year}
{2009})}\BibitemShut {NoStop}%
\bibitem [{\citenamefont {Farano}\ \emph {et~al.}(2018)\citenamefont {Farano},
\citenamefont {Cherubini}, \citenamefont {Robinet}, \citenamefont
{De~Palma},\ and\ \citenamefont {Schneider}}]{farano_2018a}%
\BibitemOpen
\bibfield  {author} {\bibinfo {author} {\bibfnamefont {M.}~\bibnamefont
{Farano}}, \bibinfo {author} {\bibfnamefont {S.}~\bibnamefont {Cherubini}},
\bibinfo {author} {\bibfnamefont {J.-C.}\ \bibnamefont {Robinet}}, \bibinfo
{author} {\bibfnamefont {P.}~\bibnamefont {De~Palma}}, \ and\ \bibinfo
{author} {\bibfnamefont {T.~M.}\ \bibnamefont {Schneider}},\ }\href {\doibase
10.1017/jfm.2018.860} {\bibfield  {journal} {\bibinfo  {journal} {Journal of
Fluid Mechanics}\ }\textbf {\bibinfo {volume} {858}},\ \bibinfo {pages} {R3}
(\bibinfo {year} {2018})}\BibitemShut {NoStop}%
\bibitem [{\citenamefont {Kerswell}\ and\ \citenamefont
{Tutty}(2007)}]{kerswell_2007}%
\BibitemOpen
\bibfield  {author} {\bibinfo {author} {\bibfnamefont {R.~R.}\ \bibnamefont
{Kerswell}}\ and\ \bibinfo {author} {\bibfnamefont {O.~R.}\ \bibnamefont
{Tutty}},\ }\href {\doibase 10.1017/S0022112007006301} {\bibfield  {journal}
{\bibinfo  {journal} {J. Fluid Mech.}\ }\textbf {\bibinfo {volume} {584}},\
\bibinfo {pages} {69–102} (\bibinfo {year} {2007})}\BibitemShut {NoStop}%
\bibitem [{\citenamefont {Kawahara}\ and\ \citenamefont
{Kida}(2001)}]{kawahara_2001}%
\BibitemOpen
\bibfield  {author} {\bibinfo {author} {\bibfnamefont {G.}~\bibnamefont
{Kawahara}}\ and\ \bibinfo {author} {\bibfnamefont {S.}~\bibnamefont
{Kida}},\ }\href@noop {} {\bibfield  {journal} {\bibinfo  {journal} {J. Fluid
Mech.}\ }\textbf {\bibinfo {volume} {449}},\ \bibinfo {pages} {291} (\bibinfo
{year} {2001})}\BibitemShut {NoStop}%
\bibitem [{\citenamefont {Kawahara}(2005)}]{kawahara_2005}%
\BibitemOpen
\bibfield  {author} {\bibinfo {author} {\bibfnamefont {G.}~\bibnamefont
{Kawahara}},\ }\href@noop {} {\bibfield  {journal} {\bibinfo  {journal}
{Phys. Fluids}\ }\textbf {\bibinfo {volume} {17}},\ \bibinfo {pages} {041702}
(\bibinfo {year} {2005})}\BibitemShut {NoStop}%
\bibitem [{\citenamefont {Viswanath}(2007)}]{viswanath_2007}%
\BibitemOpen
\bibfield  {author} {\bibinfo {author} {\bibfnamefont {D.}~\bibnamefont
{Viswanath}},\ }\href@noop {} {\bibfield  {journal} {\bibinfo  {journal} {J.
Fluid Mech.}\ }\textbf {\bibinfo {volume} {580}},\ \bibinfo {pages} {339}
(\bibinfo {year} {2007})}\BibitemShut {NoStop}%
\bibitem [{\citenamefont {Gibson}\ \emph {et~al.}(2009)\citenamefont {Gibson},
\citenamefont {Halcrow},\ and\ \citenamefont {Cvitanovi{\'c}}}]{gibson_2009}%
\BibitemOpen
\bibfield  {author} {\bibinfo {author} {\bibfnamefont {J.~F.}\ \bibnamefont
{Gibson}}, \bibinfo {author} {\bibfnamefont {J.}~\bibnamefont {Halcrow}}, \
and\ \bibinfo {author} {\bibfnamefont {P.}~\bibnamefont {Cvitanovi{\'c}}},\
}\href@noop {} {\bibfield  {journal} {\bibinfo  {journal} {J. Fluid Mech.}\
}\textbf {\bibinfo {volume} {638}},\ \bibinfo {pages} {243} (\bibinfo {year}
{2009})}\BibitemShut {NoStop}%
\bibitem [{\citenamefont {Faisst}\ and\ \citenamefont
{Eckhardt}(2003)}]{faisst_2003}%
\BibitemOpen
\bibfield  {author} {\bibinfo {author} {\bibfnamefont {H.}~\bibnamefont
{Faisst}}\ and\ \bibinfo {author} {\bibfnamefont {B.}~\bibnamefont
{Eckhardt}},\ }\href@noop {} {\bibfield  {journal} {\bibinfo  {journal}
{Phys. Rev. Lett.}\ }\textbf {\bibinfo {volume} {91}},\ \bibinfo {pages}
{224502} (\bibinfo {year} {2003})}\BibitemShut {NoStop}%
\bibitem [{\citenamefont {Wedin}\ and\ \citenamefont
{Kerswell}(2004)}]{wedin_2004}%
\BibitemOpen
\bibfield  {author} {\bibinfo {author} {\bibfnamefont {H.}~\bibnamefont
{Wedin}}\ and\ \bibinfo {author} {\bibfnamefont {R.}~\bibnamefont
{Kerswell}},\ }\href@noop {} {\bibfield  {journal} {\bibinfo  {journal} {J.
Fluid Mech.}\ }\textbf {\bibinfo {volume} {508}},\ \bibinfo {pages} {333}
(\bibinfo {year} {2004})}\BibitemShut {NoStop}%
\bibitem [{\citenamefont {Pringle}\ and\ \citenamefont
{Kerswell}(2007)}]{pringle_2007}%
\BibitemOpen
\bibfield  {author} {\bibinfo {author} {\bibfnamefont {C.~C.~T.}\
\bibnamefont {Pringle}}\ and\ \bibinfo {author} {\bibfnamefont {R.~R.}\
\bibnamefont {Kerswell}},\ }\href {\doibase 10.1103/PhysRevLett.99.074502}
{\bibfield  {journal} {\bibinfo  {journal} {Phys. Rev. Lett.}\ }\textbf
{\bibinfo {volume} {99}},\ \bibinfo {pages} {074502} (\bibinfo {year}
{2007})}\BibitemShut {NoStop}%
\bibitem [{\citenamefont {Waleffe}(2001)}]{waleffe_2001}%
\BibitemOpen
\bibfield  {author} {\bibinfo {author} {\bibfnamefont {F.}~\bibnamefont
{Waleffe}},\ }\href@noop {} {\bibfield  {journal} {\bibinfo  {journal} {J.
Fluid Mech.}\ }\textbf {\bibinfo {volume} {435}},\ \bibinfo {pages} {93}
(\bibinfo {year} {2001})}\BibitemShut {NoStop}%
\bibitem [{\citenamefont {Itano}\ and\ \citenamefont {Toh}(2001)}]{itano_2001}%
\BibitemOpen
\bibfield  {author} {\bibinfo {author} {\bibfnamefont {T.}~\bibnamefont
{Itano}}\ and\ \bibinfo {author} {\bibfnamefont {S.}~\bibnamefont {Toh}},\
}\href@noop {} {\bibfield  {journal} {\bibinfo  {journal} {J. Phys. Soc.
Jpn.}\ }\textbf {\bibinfo {volume} {70}},\ \bibinfo {pages} {703} (\bibinfo
{year} {2001})}\BibitemShut {NoStop}%
\bibitem [{\citenamefont {Toh}\ and\ \citenamefont {Itano}(2003)}]{toh_2003}%
\BibitemOpen
\bibfield  {author} {\bibinfo {author} {\bibfnamefont {S.}~\bibnamefont
{Toh}}\ and\ \bibinfo {author} {\bibfnamefont {T.}~\bibnamefont {Itano}},\
}\href {\doibase 10.1017/S0022112003003768} {\bibfield  {journal} {\bibinfo
{journal} {J. Fluid Mech.}\ }\textbf {\bibinfo {volume} {481}},\ \bibinfo
{pages} {67–76} (\bibinfo {year} {2003})}\BibitemShut {NoStop}%
\bibitem [{\citenamefont {Kim}\ \emph {et~al.}(1987)\citenamefont {Kim},
\citenamefont {Moin},\ and\ \citenamefont {Moser}}]{kim_1987}%
\BibitemOpen
\bibfield  {author} {\bibinfo {author} {\bibfnamefont {J.}~\bibnamefont
{Kim}}, \bibinfo {author} {\bibfnamefont {P.}~\bibnamefont {Moin}}, \ and\
\bibinfo {author} {\bibfnamefont {R.}~\bibnamefont {Moser}},\ }\href
{\doibase 10.1017/S0022112087000892} {\bibfield  {journal} {\bibinfo
{journal} {Journal of Fluid Mechanics}\ }\textbf {\bibinfo {volume} {177}},\
\bibinfo {pages} {133} (\bibinfo {year} {1987})}\BibitemShut {NoStop}%
\bibitem [{\citenamefont {Hamilton}\ \emph {et~al.}(1995)\citenamefont
{Hamilton}, \citenamefont {Kim},\ and\ \citenamefont
{Waleffe}}]{hamilton_1995}%
\BibitemOpen
\bibfield  {author} {\bibinfo {author} {\bibfnamefont {J.~M.}\ \bibnamefont
{Hamilton}}, \bibinfo {author} {\bibfnamefont {J.}~\bibnamefont {Kim}}, \
and\ \bibinfo {author} {\bibfnamefont {F.}~\bibnamefont {Waleffe}},\
}\href@noop {} {\bibfield  {journal} {\bibinfo  {journal} {J. Fluid Mech.}\
}\textbf {\bibinfo {volume} {287}} (\bibinfo {year} {1995})}\BibitemShut
{NoStop}%
\bibitem [{\citenamefont {Wang}\ \emph {et~al.}(2007)\citenamefont {Wang},
\citenamefont {Gibson},\ and\ \citenamefont {Waleffe}}]{wang_2007}%
\BibitemOpen
\bibfield  {author} {\bibinfo {author} {\bibfnamefont {J.}~\bibnamefont
{Wang}}, \bibinfo {author} {\bibfnamefont {J.}~\bibnamefont {Gibson}}, \ and\
\bibinfo {author} {\bibfnamefont {F.}~\bibnamefont {Waleffe}},\ }\href
{\doibase 10.1103/PhysRevLett.98.204501} {\bibfield  {journal} {\bibinfo
{journal} {Phys. Rev. Lett.}\ }\textbf {\bibinfo {volume} {98}},\ \bibinfo
{pages} {204501} (\bibinfo {year} {2007})}\BibitemShut {NoStop}%
\bibitem [{\citenamefont {Duguet}\ \emph
{et~al.}(2008{\natexlab{b}})\citenamefont {Duguet}, \citenamefont {Pringle},\
and\ \citenamefont {Kerswell}}]{duguet_2008a}%
\BibitemOpen
\bibfield  {author} {\bibinfo {author} {\bibfnamefont {Y.}~\bibnamefont
{Duguet}}, \bibinfo {author} {\bibfnamefont {C.~C.~T.}\ \bibnamefont
{Pringle}}, \ and\ \bibinfo {author} {\bibfnamefont {R.~R.}\ \bibnamefont
{Kerswell}},\ }\href@noop {} {\bibfield  {journal} {\bibinfo  {journal}
{Phys. Fluids}\ }\textbf {\bibinfo {volume} {20}},\ \bibinfo {pages} {114102}
(\bibinfo {year} {2008}{\natexlab{b}})}\BibitemShut {NoStop}%
\bibitem [{\citenamefont {Cvitanovi{\'c}}\ and\ \citenamefont
{Gibson}(2010)}]{cvitanovic_2010}%
\BibitemOpen
\bibfield  {author} {\bibinfo {author} {\bibfnamefont {P.}~\bibnamefont
{Cvitanovi{\'c}}}\ and\ \bibinfo {author} {\bibfnamefont {J.~F.}\
\bibnamefont {Gibson}},\ }\href@noop {} {\bibfield  {journal} {\bibinfo
{journal} {Phys. Scripta}\ }\textbf {\bibinfo {volume} {2010}},\ \bibinfo
{pages} {014007} (\bibinfo {year} {2010})}\BibitemShut {NoStop}%
\bibitem [{\citenamefont {Budanur}\ \emph {et~al.}(2017)\citenamefont
{Budanur}, \citenamefont {Short}, \citenamefont {Farazmand}, \citenamefont
{Willis},\ and\ \citenamefont {Cvitanović}}]{budanur_2017a}%
\BibitemOpen
\bibfield  {author} {\bibinfo {author} {\bibfnamefont {N.~B.}\ \bibnamefont
{Budanur}}, \bibinfo {author} {\bibfnamefont {K.~Y.}\ \bibnamefont {Short}},
\bibinfo {author} {\bibfnamefont {M.}~\bibnamefont {Farazmand}}, \bibinfo
{author} {\bibfnamefont {A.~P.}\ \bibnamefont {Willis}}, \ and\ \bibinfo
{author} {\bibfnamefont {P.}~\bibnamefont {Cvitanović}},\ }\href {\doibase
10.1017/jfm.2017.699} {\bibfield  {journal} {\bibinfo  {journal} {J. Fluid
Mech.}\ }\textbf {\bibinfo {volume} {833}},\ \bibinfo {pages} {274–301}
(\bibinfo {year} {2017})}\BibitemShut {NoStop}%
\bibitem [{\citenamefont {Lemoult}\ \emph {et~al.}(2014)\citenamefont
{Lemoult}, \citenamefont {Gumowski}, \citenamefont {Aider},\ and\
\citenamefont {Wesfreid}}]{lemoult_2014}%
\BibitemOpen
\bibfield  {author} {\bibinfo {author} {\bibfnamefont {G.}~\bibnamefont
{Lemoult}}, \bibinfo {author} {\bibfnamefont {K.}~\bibnamefont {Gumowski}},
\bibinfo {author} {\bibfnamefont {J.-L.}\ \bibnamefont {Aider}}, \ and\
\bibinfo {author} {\bibfnamefont {J.~E.}\ \bibnamefont {Wesfreid}},\ }\href
{\doibase 10.1140/epje/i2014-14025-2} {\bibfield  {journal} {\bibinfo
{journal} {Eur. Phys. J. E}\ }\textbf {\bibinfo {volume} {37}},\ \bibinfo
{pages} {1} (\bibinfo {year} {2014})}\BibitemShut {NoStop}%
\bibitem [{\citenamefont {Budanur}\ and\ \citenamefont
{Hof}(2017)}]{budanur_2017b}%
\BibitemOpen
\bibfield  {author} {\bibinfo {author} {\bibfnamefont {N.~B.}\ \bibnamefont
{Budanur}}\ and\ \bibinfo {author} {\bibfnamefont {B.}~\bibnamefont {Hof}},\
}\href {\doibase 10.1017/jfm.2017.516} {\bibfield  {journal} {\bibinfo
{journal} {Journal of Fluid Mechanics}\ }\textbf {\bibinfo {volume} {827}},\
\bibinfo {pages} {R1} (\bibinfo {year} {2017})}\BibitemShut {NoStop}%
\bibitem [{\citenamefont {Smale}(1967)}]{smale_1967}%
\BibitemOpen
\bibfield  {author} {\bibinfo {author} {\bibfnamefont {S.}~\bibnamefont
{Smale}},\ }\href@noop {} {\bibfield  {journal} {\bibinfo  {journal}
{Bulletin of the American mathematical Society}\ }\textbf {\bibinfo {volume}
{73}},\ \bibinfo {pages} {747} (\bibinfo {year} {1967})}\BibitemShut
{NoStop}%
\bibitem [{\citenamefont {van Veen}\ and\ \citenamefont
{Kawahara}(2011)}]{vanveen_2011b}%
\BibitemOpen
\bibfield  {author} {\bibinfo {author} {\bibfnamefont {L.}~\bibnamefont {van
Veen}}\ and\ \bibinfo {author} {\bibfnamefont {G.}~\bibnamefont {Kawahara}},\
}\href {\doibase 10.1103/PhysRevLett.107.114501} {\bibfield  {journal}
{\bibinfo  {journal} {Phys. Rev. Lett.}\ }\textbf {\bibinfo {volume} {107}},\
\bibinfo {pages} {114501} (\bibinfo {year} {2011})}\BibitemShut {NoStop}%
\bibitem [{\citenamefont {van Veen}\ \emph {et~al.}(2011)\citenamefont {van
Veen}, \citenamefont {Kawahara},\ and\ \citenamefont
{Atsushi}}]{vanveen_2011a}%
\BibitemOpen
\bibfield  {author} {\bibinfo {author} {\bibfnamefont {L.}~\bibnamefont {van
Veen}}, \bibinfo {author} {\bibfnamefont {G.}~\bibnamefont {Kawahara}}, \
and\ \bibinfo {author} {\bibfnamefont {M.}~\bibnamefont {Atsushi}},\ }\href
{\doibase 10.1137/100789804} {\bibfield  {journal} {\bibinfo  {journal} {SIAM
J. Sci. Comput.}\ }\textbf {\bibinfo {volume} {33}},\ \bibinfo {pages} {25}
(\bibinfo {year} {2011})}\BibitemShut {NoStop}%
\bibitem [{\citenamefont {Kline}\ \emph {et~al.}(1967)\citenamefont {Kline},
\citenamefont {Reynolds}, \citenamefont {Schraub},\ and\ \citenamefont
{Runstadler}}]{kline_1967}%
\BibitemOpen
\bibfield  {author} {\bibinfo {author} {\bibfnamefont {S.}~\bibnamefont
{Kline}}, \bibinfo {author} {\bibfnamefont {W.}~\bibnamefont {Reynolds}},
\bibinfo {author} {\bibfnamefont {F.}~\bibnamefont {Schraub}}, \ and\
\bibinfo {author} {\bibfnamefont {P.}~\bibnamefont {Runstadler}},\
}\href@noop {} {\bibfield  {journal} {\bibinfo  {journal} {J. Fluid Mech.}\
}\textbf {\bibinfo {volume} {30}},\ \bibinfo {pages} {741} (\bibinfo {year}
{1967})}\BibitemShut {NoStop}%
\bibitem [{\citenamefont {Riols}\ \emph {et~al.}(2013)\citenamefont {Riols},
\citenamefont {Rincon}, \citenamefont {Cossu}, \citenamefont {Lesur},
\citenamefont {Longaretti}, \citenamefont {Ogilvie},\ and\ \citenamefont
{Herault}}]{riols_2013}%
\BibitemOpen
\bibfield  {author} {\bibinfo {author} {\bibfnamefont {A.}~\bibnamefont
{Riols}}, \bibinfo {author} {\bibfnamefont {F.}~\bibnamefont {Rincon}},
\bibinfo {author} {\bibfnamefont {C.}~\bibnamefont {Cossu}}, \bibinfo
{author} {\bibfnamefont {G.}~\bibnamefont {Lesur}}, \bibinfo {author}
{\bibfnamefont {P.-Y.}\ \bibnamefont {Longaretti}}, \bibinfo {author}
{\bibfnamefont {G.~I.}\ \bibnamefont {Ogilvie}}, \ and\ \bibinfo {author}
{\bibfnamefont {J.}~\bibnamefont {Herault}},\ }\href {\doibase
10.1017/jfm.2013.317} {\bibfield  {journal} {\bibinfo  {journal} {Journal of
Fluid Mechanics}\ }\textbf {\bibinfo {volume} {731}},\ \bibinfo {pages} {1}
(\bibinfo {year} {2013})}\BibitemShut {NoStop}%
\bibitem [{\citenamefont {Pershin}\ \emph {et~al.}(2019)\citenamefont
{Pershin}, \citenamefont {Beaume},\ and\ \citenamefont
{Tobias}}]{pershin_2019}%
\BibitemOpen
\bibfield  {author} {\bibinfo {author} {\bibfnamefont {A.}~\bibnamefont
{Pershin}}, \bibinfo {author} {\bibfnamefont {C.}~\bibnamefont {Beaume}}, \
and\ \bibinfo {author} {\bibfnamefont {S.~M.}\ \bibnamefont {Tobias}},\
}\href {\doibase 10.1017/jfm.2019.154} {\bibfield  {journal} {\bibinfo
{journal} {Journal of Fluid Mechanics}\ }\textbf {\bibinfo {volume} {867}},\
\bibinfo {pages} {414} (\bibinfo {year} {2019})}\BibitemShut {NoStop}%
\bibitem [{\citenamefont {{Burak Budanur}}\ \emph {et~al.}(2018)\citenamefont
{{Burak Budanur}}, \citenamefont {{Shaurya Dogra}},\ and\ \citenamefont
{{Hof}}}]{budanur_2019}%
\BibitemOpen
\bibfield  {author} {\bibinfo {author} {\bibfnamefont {N.}~\bibnamefont
{{Burak Budanur}}}, \bibinfo {author} {\bibfnamefont {A.}~\bibnamefont
{{Shaurya Dogra}}}, \ and\ \bibinfo {author} {\bibfnamefont {B.}~\bibnamefont
{{Hof}}},\ }\href@noop {} {\bibfield  {journal} {\bibinfo  {journal} {arXiv
e-prints}\ ,\ \bibinfo {eid} {arXiv:1810.02211}} (\bibinfo {year} {2018})},\
\Eprint {http://arxiv.org/abs/1810.02211} {arXiv:1810.02211
[physics.flu-dyn]} \BibitemShut {NoStop}%
\bibitem [{\citenamefont {Bondarenko}\ \emph {et~al.}(1979)\citenamefont
{Bondarenko}, \citenamefont {Gak},\ and\ \citenamefont
{Dolzhanskiy}}]{bondarenko_1979}%
\BibitemOpen
\bibfield  {author} {\bibinfo {author} {\bibfnamefont {N.~F.}\ \bibnamefont
{Bondarenko}}, \bibinfo {author} {\bibfnamefont {M.~Z.}\ \bibnamefont {Gak}},
\ and\ \bibinfo {author} {\bibfnamefont {F.~V.}\ \bibnamefont
{Dolzhanskiy}},\ }\href@noop {} {\bibfield  {journal} {\bibinfo  {journal}
{Izv. Akad. Nauk SSSR, Fiz. Atmos. Okeana}\ }\textbf {\bibinfo {volume}
{15}},\ \bibinfo {pages} {711} (\bibinfo {year} {1979})}\BibitemShut
{NoStop}%
\bibitem [{\citenamefont {Suri}\ \emph {et~al.}(2014)\citenamefont {Suri},
\citenamefont {Tithof}, \citenamefont {Mitchell}, \citenamefont {Grigoriev},\
and\ \citenamefont {Schatz}}]{suri_2014}%
\BibitemOpen
\bibfield  {author} {\bibinfo {author} {\bibfnamefont {B.}~\bibnamefont
{Suri}}, \bibinfo {author} {\bibfnamefont {J.}~\bibnamefont {Tithof}},
\bibinfo {author} {\bibfnamefont {R.}~\bibnamefont {Mitchell}}, \bibinfo
{author} {\bibfnamefont {R.~O.}\ \bibnamefont {Grigoriev}}, \ and\ \bibinfo
{author} {\bibfnamefont {M.~F.}\ \bibnamefont {Schatz}},\ }\href {\doibase
http://dx.doi.org/10.1063/1.4873417} {\bibfield  {journal} {\bibinfo
{journal} {Phys. Fluids}\ }\textbf {\bibinfo {volume} {26}},\ \bibinfo {eid}
{053601} (\bibinfo {year} {2014})}\BibitemShut {NoStop}%
\bibitem [{\citenamefont {Chandler}\ and\ \citenamefont
{Kerswell}(2013)}]{chandler_2013}%
\BibitemOpen
\bibfield  {author} {\bibinfo {author} {\bibfnamefont {G.~J.}\ \bibnamefont
{Chandler}}\ and\ \bibinfo {author} {\bibfnamefont {R.~R.}\ \bibnamefont
{Kerswell}},\ }\href@noop {} {\bibfield  {journal} {\bibinfo  {journal} {J.
Fluid Mech.}\ }\textbf {\bibinfo {volume} {722}},\ \bibinfo {pages} {554}
(\bibinfo {year} {2013})}\BibitemShut {NoStop}%
\bibitem [{\citenamefont {Lucas}\ and\ \citenamefont
{Kerswell}(2014)}]{lucas_2014}%
\BibitemOpen
\bibfield  {author} {\bibinfo {author} {\bibfnamefont {D.}~\bibnamefont
{Lucas}}\ and\ \bibinfo {author} {\bibfnamefont {R.~R.}\ \bibnamefont
{Kerswell}},\ }\href@noop {} {\bibfield  {journal} {\bibinfo  {journal} {J.
Fluid Mech.}\ }\textbf {\bibinfo {volume} {750}},\ \bibinfo {pages} {518}
(\bibinfo {year} {2014})}\BibitemShut {NoStop}%
\bibitem [{\citenamefont {Farazmand}(2016)}]{farazmand_2016}%
\BibitemOpen
\bibfield  {author} {\bibinfo {author} {\bibfnamefont {M.}~\bibnamefont
{Farazmand}},\ }\href {\doibase 10.1017/jfm.2016.203} {\bibfield  {journal}
{\bibinfo  {journal} {J. Fluid Mech.}\ }\textbf {\bibinfo {volume} {795}},\
\bibinfo {pages} {278–312} (\bibinfo {year} {2016})}\BibitemShut {NoStop}%
\bibitem [{\citenamefont {Cvitanovi\ifmmode~\acute{c}\else
\'{c}\fi{}}(1988)}]{cvitanovic_1988}%
\BibitemOpen
\bibfield  {author} {\bibinfo {author} {\bibfnamefont {P.}~\bibnamefont
{Cvitanovi\ifmmode~\acute{c}\else \'{c}\fi{}}},\ }\href {\doibase
10.1103/PhysRevLett.61.2729} {\bibfield  {journal} {\bibinfo  {journal}
{Phys. Rev. Lett.}\ }\textbf {\bibinfo {volume} {61}},\ \bibinfo {pages}
{2729} (\bibinfo {year} {1988})}\BibitemShut {NoStop}%
\bibitem [{\citenamefont {Tithof}\ \emph {et~al.}(2017)\citenamefont {Tithof},
\citenamefont {Suri}, \citenamefont {Pallantla}, \citenamefont {Grigoriev},\
and\ \citenamefont {Schatz}}]{tithof_2017}%
\BibitemOpen
\bibfield  {author} {\bibinfo {author} {\bibfnamefont {J.}~\bibnamefont
{Tithof}}, \bibinfo {author} {\bibfnamefont {B.}~\bibnamefont {Suri}},
\bibinfo {author} {\bibfnamefont {R.~K.}\ \bibnamefont {Pallantla}}, \bibinfo
{author} {\bibfnamefont {R.~O.}\ \bibnamefont {Grigoriev}}, \ and\ \bibinfo
{author} {\bibfnamefont {M.~F.}\ \bibnamefont {Schatz}},\ }\href@noop {}
{\bibfield  {journal} {\bibinfo  {journal} {J. Fluid Mech.}\ }\textbf
{\bibinfo {volume} {828}},\ \bibinfo {pages} {837} (\bibinfo {year}
{2017})}\BibitemShut {NoStop}%
\bibitem [{\citenamefont {Armfield}\ and\ \citenamefont
{Street}(1999)}]{armfield_1999}%
\BibitemOpen
\bibfield  {author} {\bibinfo {author} {\bibfnamefont {S.}~\bibnamefont
{Armfield}}\ and\ \bibinfo {author} {\bibfnamefont {R.}~\bibnamefont
{Street}},\ }\href {\doibase http://dx.doi.org/10.1006/jcph.1999.6275}
{\bibfield  {journal} {\bibinfo  {journal} {J. Comput. Phys.}\ }\textbf
{\bibinfo {volume} {153}},\ \bibinfo {pages} {660 } (\bibinfo {year}
{1999})}\BibitemShut {NoStop}%
\bibitem [{\citenamefont {Suri}(2017)}]{suri_2017b}%
\BibitemOpen
\bibfield  {author} {\bibinfo {author} {\bibfnamefont {B.}~\bibnamefont
{Suri}},\ }\emph {\bibinfo {title} {Quasi-Two-Dimensional Kolmogorov flow:
Bifurcations and Exact Coherent Structures}},\ \href@noop {} {Ph.D. thesis},\
\bibinfo  {school} {Georgia Institute of Technology} (\bibinfo {year}
{2017})\BibitemShut {NoStop}%
\bibitem [{\citenamefont {Farmer}\ \emph {et~al.}(1983)\citenamefont {Farmer},
\citenamefont {Ott},\ and\ \citenamefont {Yorke}}]{farmer_1983}%
\BibitemOpen
\bibfield  {author} {\bibinfo {author} {\bibfnamefont {J.}~\bibnamefont
{Farmer}}, \bibinfo {author} {\bibfnamefont {E.}~\bibnamefont {Ott}}, \ and\
\bibinfo {author} {\bibfnamefont {J.~A.}\ \bibnamefont {Yorke}},\ }\href
{\doibase https://doi.org/10.1016/0167-2789(83)90125-2} {\bibfield  {journal}
{\bibinfo  {journal} {Physica D: Nonlinear Phenomena}\ }\textbf {\bibinfo
{volume} {7}},\ \bibinfo {pages} {153 } (\bibinfo {year} {1983})}\BibitemShut
{NoStop}%
\bibitem [{\citenamefont {Neelavara}\ \emph {et~al.}(2017)\citenamefont
{Neelavara}, \citenamefont {Duguet},\ and\ \citenamefont
{Lusseyran}}]{acharya_2017}%
\BibitemOpen
\bibfield  {author} {\bibinfo {author} {\bibfnamefont {S.~A.}\ \bibnamefont
{Neelavara}}, \bibinfo {author} {\bibfnamefont {Y.}~\bibnamefont {Duguet}}, \
and\ \bibinfo {author} {\bibfnamefont {F.}~\bibnamefont {Lusseyran}},\
}\href@noop {} {\bibfield  {journal} {\bibinfo  {journal} {Fluid Dynamics
Research}\ }\textbf {\bibinfo {volume} {49}},\ \bibinfo {pages} {035511}
(\bibinfo {year} {2017})}\BibitemShut {NoStop}%
\bibitem [{\citenamefont {Saad}\ and\ \citenamefont
{Schultz}(1986)}]{saad_1986}%
\BibitemOpen
\bibfield  {author} {\bibinfo {author} {\bibfnamefont {Y.}~\bibnamefont
{Saad}}\ and\ \bibinfo {author} {\bibfnamefont {M.~H.}\ \bibnamefont
{Schultz}},\ }\href@noop {} {\bibfield  {journal} {\bibinfo  {journal} {SIAM
J. Sci. Stat. Comp.}\ }\textbf {\bibinfo {volume} {7}},\ \bibinfo {pages}
{856} (\bibinfo {year} {1986})}\BibitemShut {NoStop}%
\bibitem [{\citenamefont {Kelley}(2003)}]{kelley_2003}%
\BibitemOpen
\bibfield  {author} {\bibinfo {author} {\bibfnamefont {C.}~\bibnamefont
{Kelley}},\ }\href {\doibase 10.1137/1.9780898718898} {\emph {\bibinfo
{title} {Solving Nonlinear Equations with Newton's Method}}}\ (\bibinfo
{publisher} {SIAM},\ \bibinfo {year} {2003})\BibitemShut {NoStop}%
\bibitem [{\citenamefont {Mitchell}(2013)}]{mitchell_2013}%
\BibitemOpen
\bibfield  {author} {\bibinfo {author} {\bibfnamefont {R.}~\bibnamefont
{Mitchell}},\ }\emph {\bibinfo {title} {Transition to turbulence and mixing
in a quasi-two-dimensional {L}orentz force-driven {K}olmogorov flow}},\
\href@noop {} {Ph.D. thesis},\ \bibinfo  {school} {Georgia Institute of
Technology} (\bibinfo {year} {2013})\BibitemShut {NoStop}%
\bibitem [{\citenamefont {Avila}\ \emph {et~al.}(2013)\citenamefont {Avila},
\citenamefont {Mellibovsky}, \citenamefont {Roland},\ and\ \citenamefont
{Hof}}]{avila_2013}%
\BibitemOpen
\bibfield  {author} {\bibinfo {author} {\bibfnamefont {M.}~\bibnamefont
{Avila}}, \bibinfo {author} {\bibfnamefont {F.}~\bibnamefont {Mellibovsky}},
\bibinfo {author} {\bibfnamefont {N.}~\bibnamefont {Roland}}, \ and\ \bibinfo
{author} {\bibfnamefont {B.}~\bibnamefont {Hof}},\ }\href {\doibase
10.1103/PhysRevLett.110.224502} {\bibfield  {journal} {\bibinfo  {journal}
{Phys. Rev. Lett.}\ }\textbf {\bibinfo {volume} {110}},\ \bibinfo {pages}
{224502} (\bibinfo {year} {2013})}\BibitemShut {NoStop}%
\bibitem [{\citenamefont {Hof}\ \emph {et~al.}(2006)\citenamefont {Hof},
\citenamefont {Westerweel}, \citenamefont {Schneider},\ and\ \citenamefont
{Eckhardt}}]{hof_2006}%
\BibitemOpen
\bibfield  {author} {\bibinfo {author} {\bibfnamefont {B.}~\bibnamefont
{Hof}}, \bibinfo {author} {\bibfnamefont {J.}~\bibnamefont {Westerweel}},
\bibinfo {author} {\bibfnamefont {T.~M.}\ \bibnamefont {Schneider}}, \ and\
\bibinfo {author} {\bibfnamefont {B.}~\bibnamefont {Eckhardt}},\ }\href@noop
{} {\bibfield  {journal} {\bibinfo  {journal} {Nature}\ }\textbf {\bibinfo
{volume} {443}},\ \bibinfo {pages} {59} (\bibinfo {year} {2006})}\BibitemShut
{NoStop}%
\bibitem [{\citenamefont {Eckhardt}\ \emph {et~al.}(2007)\citenamefont
{Eckhardt}, \citenamefont {Faisst}, \citenamefont {Schmiegel},\ and\
\citenamefont {Schneider}}]{eckhardt2007dynamical}%
\BibitemOpen
\bibfield  {author} {\bibinfo {author} {\bibfnamefont {B.}~\bibnamefont
{Eckhardt}}, \bibinfo {author} {\bibfnamefont {H.}~\bibnamefont {Faisst}},
\bibinfo {author} {\bibfnamefont {A.}~\bibnamefont {Schmiegel}}, \ and\
\bibinfo {author} {\bibfnamefont {T.~M.}\ \bibnamefont {Schneider}},\
}\href@noop {} {\bibfield  {journal} {\bibinfo  {journal} {Philosophical
Transactions of the Royal Society A: Mathematical, Physical and Engineering
Sciences}\ }\textbf {\bibinfo {volume} {366}},\ \bibinfo {pages} {1297}
(\bibinfo {year} {2007})}\BibitemShut {NoStop}%
\bibitem [{\citenamefont {Schneider}\ \emph {et~al.}(2010)\citenamefont
{Schneider}, \citenamefont {De~Lillo}, \citenamefont {Buehrle}, \citenamefont
{Eckhardt}, \citenamefont {D{\"o}rnemann}, \citenamefont {D{\"o}rnemann},\
and\ \citenamefont {Freisleben}}]{schneider2010transient}%
\BibitemOpen
\bibfield  {author} {\bibinfo {author} {\bibfnamefont {T.~M.}\ \bibnamefont
{Schneider}}, \bibinfo {author} {\bibfnamefont {F.}~\bibnamefont {De~Lillo}},
\bibinfo {author} {\bibfnamefont {J.}~\bibnamefont {Buehrle}}, \bibinfo
{author} {\bibfnamefont {B.}~\bibnamefont {Eckhardt}}, \bibinfo {author}
{\bibfnamefont {T.}~\bibnamefont {D{\"o}rnemann}}, \bibinfo {author}
{\bibfnamefont {K.}~\bibnamefont {D{\"o}rnemann}}, \ and\ \bibinfo {author}
{\bibfnamefont {B.}~\bibnamefont {Freisleben}},\ }\href@noop {} {\bibfield
{journal} {\bibinfo  {journal} {Physical review E}\ }\textbf {\bibinfo
{volume} {81}},\ \bibinfo {pages} {015301} (\bibinfo {year}
{2010})}\BibitemShut {NoStop}%
\bibitem [{\citenamefont {Borrero-Echeverry}\ \emph {et~al.}(2010)\citenamefont
{Borrero-Echeverry}, \citenamefont {Schatz},\ and\ \citenamefont
{Tagg}}]{borrero_2010}%
\BibitemOpen
\bibfield  {author} {\bibinfo {author} {\bibfnamefont {D.}~\bibnamefont
{Borrero-Echeverry}}, \bibinfo {author} {\bibfnamefont {M.~F.}\ \bibnamefont
{Schatz}}, \ and\ \bibinfo {author} {\bibfnamefont {R.}~\bibnamefont
{Tagg}},\ }\href@noop {} {\bibfield  {journal} {\bibinfo  {journal} {Physical
Review E}\ }\textbf {\bibinfo {volume} {81}},\ \bibinfo {pages} {025301}
(\bibinfo {year} {2010})}\BibitemShut {NoStop}%
\bibitem [{\citenamefont {Kreilos}\ and\ \citenamefont
{Eckhardt}(2012)}]{kreilos_2012}%
\BibitemOpen
\bibfield  {author} {\bibinfo {author} {\bibfnamefont {T.}~\bibnamefont
{Kreilos}}\ and\ \bibinfo {author} {\bibfnamefont {B.}~\bibnamefont
{Eckhardt}},\ }\href@noop {} {\bibfield  {journal} {\bibinfo  {journal}
{Chaos: An Interdisciplinary Journal of Nonlinear Science}\ }\textbf
{\bibinfo {volume} {22}},\ \bibinfo {pages} {047505} (\bibinfo {year}
{2012})}\BibitemShut {NoStop}%
\bibitem [{\citenamefont {Kirk}\ and\ \citenamefont
{Silber}(1994)}]{kirk_1994}%
\BibitemOpen
\bibfield  {author} {\bibinfo {author} {\bibfnamefont {V.}~\bibnamefont
{Kirk}}\ and\ \bibinfo {author} {\bibfnamefont {M.}~\bibnamefont {Silber}},\
}\href@noop {} {\bibfield  {journal} {\bibinfo  {journal} {Nonlinearity}\
}\textbf {\bibinfo {volume} {7}},\ \bibinfo {pages} {1605} (\bibinfo {year}
{1994})}\BibitemShut {NoStop}%
\bibitem [{\citenamefont {Krupa}\ and\ \citenamefont
{Melbourne}(1995)}]{krupa_1995}%
\BibitemOpen
\bibfield  {author} {\bibinfo {author} {\bibfnamefont {M.}~\bibnamefont
{Krupa}}\ and\ \bibinfo {author} {\bibfnamefont {I.}~\bibnamefont
{Melbourne}},\ }\href@noop {} {\bibfield  {journal} {\bibinfo  {journal}
{Ergodic Theory and Dynamical Systems}\ }\textbf {\bibinfo {volume} {15}},\
\bibinfo {pages} {121} (\bibinfo {year} {1995})}\BibitemShut {NoStop}%
\bibitem [{\citenamefont {Hof}\ \emph {et~al.}(2010)\citenamefont {Hof},
\citenamefont {de~Lozar}, \citenamefont {Avila}, \citenamefont {Tu},\ and\
\citenamefont {Schneider}}]{hof_2010}%
\BibitemOpen
\bibfield  {author} {\bibinfo {author} {\bibfnamefont {B.}~\bibnamefont
{Hof}}, \bibinfo {author} {\bibfnamefont {A.}~\bibnamefont {de~Lozar}},
\bibinfo {author} {\bibfnamefont {M.}~\bibnamefont {Avila}}, \bibinfo
{author} {\bibfnamefont {X.}~\bibnamefont {Tu}}, \ and\ \bibinfo {author}
{\bibfnamefont {T.~M.}\ \bibnamefont {Schneider}},\ }\href {\doibase
10.1126/science.1186091} {\bibfield  {journal} {\bibinfo  {journal}
{Science}\ }\textbf {\bibinfo {volume} {327}},\ \bibinfo {pages} {1491}
(\bibinfo {year} {2010})}\BibitemShut {NoStop}%
\bibitem [{\citenamefont {K{\"u}hnen}\ \emph {et~al.}(2018)\citenamefont
{K{\"u}hnen}, \citenamefont {Song}, \citenamefont {Scarselli}, \citenamefont
{Budanur}, \citenamefont {Riedl}, \citenamefont {Willis}, \citenamefont
{Avila},\ and\ \citenamefont {Hof}}]{kuhnen_2018}%
\BibitemOpen
\bibfield  {author} {\bibinfo {author} {\bibfnamefont {J.}~\bibnamefont
{K{\"u}hnen}}, \bibinfo {author} {\bibfnamefont {B.}~\bibnamefont {Song}},
\bibinfo {author} {\bibfnamefont {D.}~\bibnamefont {Scarselli}}, \bibinfo
{author} {\bibfnamefont {N.~B.}\ \bibnamefont {Budanur}}, \bibinfo {author}
{\bibfnamefont {M.}~\bibnamefont {Riedl}}, \bibinfo {author} {\bibfnamefont
{A.~P.}\ \bibnamefont {Willis}}, \bibinfo {author} {\bibfnamefont
{M.}~\bibnamefont {Avila}}, \ and\ \bibinfo {author} {\bibfnamefont
{B.}~\bibnamefont {Hof}},\ }\href {\doibase 10.1038/s41567-017-0018-3}
{\bibfield  {journal} {\bibinfo  {journal} {Nature Physics}\ }\textbf
{\bibinfo {volume} {14}},\ \bibinfo {pages} {386} (\bibinfo {year}
{2018})}\BibitemShut {NoStop}%
\bibitem [{\citenamefont {Pringle}\ \emph {et~al.}(2012)\citenamefont
{Pringle}, \citenamefont {Willis},\ and\ \citenamefont
{Kerswell}}]{pringle_2012}%
\BibitemOpen
\bibfield  {author} {\bibinfo {author} {\bibfnamefont {C.~C.}\ \bibnamefont
{Pringle}}, \bibinfo {author} {\bibfnamefont {A.~P.}\ \bibnamefont {Willis}},
\ and\ \bibinfo {author} {\bibfnamefont {R.~R.}\ \bibnamefont {Kerswell}},\
}\href {\doibase 10.1017/jfm.2012.192} {\bibfield  {journal} {\bibinfo
{journal} {Journal of Fluid Mechanics}\ }\textbf {\bibinfo {volume} {702}},\
\bibinfo {pages} {415} (\bibinfo {year} {2012})}\BibitemShut {NoStop}%
\bibitem [{\citenamefont {Eckmann}\ \emph {et~al.}(1987)\citenamefont
{Eckmann}, \citenamefont {Kamphorst},\ and\ \citenamefont
{Ruelle}}]{eckmann_1987}%
\BibitemOpen
\bibfield  {author} {\bibinfo {author} {\bibfnamefont {J.-P.}\ \bibnamefont
{Eckmann}}, \bibinfo {author} {\bibfnamefont {S.~O.}\ \bibnamefont
{Kamphorst}}, \ and\ \bibinfo {author} {\bibfnamefont {D.}~\bibnamefont
{Ruelle}},\ }\href@noop {} {\bibfield  {journal} {\bibinfo  {journal}
{Europhys. Lett.}\ }\textbf {\bibinfo {volume} {4}},\ \bibinfo {pages} {973}
(\bibinfo {year} {1987})}\BibitemShut {NoStop}%
\end{thebibliography}
\end{document}